\documentstyle[seceq,epsfig]{jpsj}

\def\runtitle{
GL theory for $d$-wave pairing and fourfold symmetric vortex core structure
}

\def\runauthor{ 
Naoki {\sc Enomoto}, Masanori {\sc Ichioka} and Kazushige {\sc Machida} 
}

\title
{
Ginzburg Landau theory for $d$-wave pairing \\ 
and fourfold symmetric vortex core structure
} 

\author
{ 
Naoki {\sc Enomoto}\footnote{E-mail: enomoto@mp.okayama-u.ac.jp}, 
Masanori {\sc Ichioka}\footnote{E-mail: oka@mp.okayama-u.ac.jp} and 
Kazushige {\sc Machida}\footnote{E-mail: machida@mp.okayama-u.ac.jp}
}

\inst
{
Department of Physics, Okayama University, Okayama 
700
}

\recdate
{
July 12, 1996
}

\abst
{
The Ginzburg Landau theory for $d_{x^2-y^2}$-wave superconductors 
is constructed, by starting from the Gor'kov equation with including 
correction terms up to the next order of $\ln(T_c/T)$. 
Some of the non-local correction terms are found to break the cylindrical 
symmetry and lead to the fourfold symmetric core structure, reflecting the 
internal degree of freedom in the pair potential. 
Using this extended Ginzburg Landau theory, we investigate the fourfold 
symmetric structure of the pair potential, current and magnetic field 
around an isolated single vortex, and clarify concretely how the vortex 
core structure deviates from the cylindrical symmetry in the 
$d_{x^2-y^2}$-wave superconductors.
}

\kword
{
d-wave superconductor, vortex core structure, Ginzburg Landau theory
}

\begin{document}

\sloppy
\maketitle
\begin{full}

\section{Introduction}
\label{sec:1}

Much attention has been focused on a vortex structure in high-$T_c$ 
 superconductors.
It is expected to be clarified how the vortex structure of the 
high-$T_c$ superconductors is different from that of the conventional 
superconductors. 
By a number of experimental and theoretical investigations, it is concluded 
that the symmetry of these superconductors is most likely to be  
$d_{x^2-y^2}$-wave. 
Therefore, one of the points is how the vortex structure of $d_{x^2-y^2}$-wave 
superconductors is different from that of isotropic $s$-wave superconductors. 
The $d_{x^2-y^2}$-wave pairing, i.e., 
$\hat k_x^2 - \hat k_y^2$ in momentum space, has fourfold symmetry 
for the rotation about the $c$-axis. 
We expect that, reflecting this symmetry, the core structure of an 
isolated single vortex may break the cylindrical symmetry and show fourfold 
symmetry in $d_{x^2-y^2}$-wave superconductors.

In their experiments, Keimer {\it et al.}~\cite{Keimer} reported an oblique
lattice by a small-angle neutron scattering study of the vortex lattice
in ${\rm YBa_2Cu_3O_7}$ in a magnetic field region of 0.5T$\le H \le$5T 
applied parallel to the $c$-axis.
The vortex lattice has an angle of $73^\circ$ between the two primitive
vectors and is oriented such that the nearest-neighbor direction of vortices
makes an angle of $45^\circ$ with the $a$-axis.
The oblique lattice was also observed by scanning tunneling microscopy
(STM) by Maggio-Aprile {\it et al.}~\cite{Maggio,Renner} 
They also observed the elliptic-shaped STM image of the vortex core, 
and concluded that this oblique lattice cannot be explained by considering 
only the effect of the intrinsic in-plane anisotropy, that is,
the difference of the coherence lengths between $a$-axis and
$b$-axis directions.~\cite{Renner,Walker}  
It is suggested that this deformation from the triangular lattice in 
$d_{x^2-y^2}$-wave superconductors is due to the effect of the fourfold 
symmetric vortex core structure.~\cite{Renner,Won,Berlinsky}

The fourfold symmetric vortex core structure in $d_{x^2-y^2}$-wave 
superconductors was so far derived theoretically when the $s$-wave component 
is induced around a vortex of $d_{x^2-y^2}$-wave 
superconductivity.~\cite{Berlinsky,Ren,Xu,Soininen,IchiokaS} 
This mixing scenario was mainly studied by Berlinsky 
{\it et al.}\cite{Berlinsky} and Ren {\it et al.}~\cite{Ren,Xu} in the 
two-component Ginzburg Landau (GL) theory for $s$- and $d$-wave 
superconductivity.
According to the consideration based on the two-component GL theory,
it is possible that the $s$-wave component is coupled with the $d$-wave
component through the gradient terms.
Therefore, the $s$-wave component may be  induced when the $d$-wave
order parameter spatially varies, such as near the vortex or interface
under certain restricted conditions.\cite{Matsumoto}
The induced $s$-wave component around the vortex is fourfold symmetric.
The resulting vortex structure in $d$-wave superconductors, therefore,
exhibits fourfold symmetry.

For this scenario to be applied to high-$T_c$ superconductors, 
the amplitude of the induced $s$-wave component should be comparable 
to the already-existing main $d_{x^2-y^2}$-wave component. 
The amplitude of the induced $s$-wave component is proportional 
to the $s$-wave component $V_s$ of the pairing interaction.~\cite{IchiokaS}  
Then, in the case where $|V_s|$ is negligibly small compared with the dominant 
$d_{x^2-y^2}$-wave  pairing interaction, the induced $s$-wave order parameter 
is negligible. 
In this paper, we consider this pure $d_{x^2-y^2}$-wave case, that is, 
the case when the induced $s$-wave component can be neglected. 
In our opinion, as the first step to understanding the fourfold symmetric 
vortex core in $d_{x^2-y^2}$-wave superconductors, the pure 
$d_{x^2-y^2}$-wave case should be studied before considering 
the mixing of the induced $s$-wave component. 
In this case, we have to consider a new scenario to explain the fourfold 
symmetric vortex core structure.

When the induced $s$-wave component can be neglected, the order parameter 
reduces to be one component, i.e., only the $d_{x^2-y^2}$-wave component, 
and the GL equation reduces to the same form as that of the isotropic 
$s$-wave case within the conventional GL theory. 
Within this framework, the vortex core structure remains cylindrically  
symmetric, and the vortex lattice forms a triangular lattice, even in 
the $d_{x^2-y^2}$-wave superconductors. 
On the other hand, Ichioka {\it et al.}~\cite{IchiokaF} showed that the 
vortex core structure is fourfold symmetric in the pure $d_{x^2-y^2}$-wave 
superconductors by using the quasi-classical Eilenberger theory. 
This means that we have to modify the conventional GL theory to explain 
the fourfold symmetric vortex core structure.
For this purpose, we consider the following scenario.

Strictly speaking, the conventional GL equation is valid only near the 
transition temperature $T_c$.
Far from $T_{\rm c}$, we have to include several correction terms of 
the order $\ln(T_c/T)$, which consist of both the higher powers of the order 
parameter and higher order derivatives, i.e., the so-called nonlocal terms. 
Among them, some nonlocal terms break cylindrical symmetry in the 
$d_{x^2-y^2}$-wave pairing case. 
Therefore, these correction terms lead to the fourfold symmetry of the 
vortex structure.

So far, these terms neglected in the conventional GL theory were  
considered to explain the fourfold symmetric behavior of the upper 
critical field $H_{c2}$ in the $ab$-plane~\cite{Koike} and the 
deformation of the vortex lattice from a triangular lattice.
As for the fourfold symmetric behavior of $H_{c2}$, Takanaka and 
Kuboya~\cite{Takanaka} showed it from the GL theory with nonlocal 
correction terms, and Won and Maki~\cite{Won,Maki} from the Gor'kov equation. 
A stable vortex lattice was discussed by Won and Maki in the pure 
$d_{x^2-y^2}$-wave superconductors.~\cite{Won}  
As for the case of the isotropic $s$-wave pairing but anisotropic density 
of states at the Fermi surface, Takanaka and Nagashima~\cite{TakanakaPTP}  
derived the GL equation with including the higher order correction terms, 
and studied the anisotropy of $H_{c2}$ and a stable vortex lattice.
However, the anisotropic vortex core structure has not been studied so far. 
In this paper, we discuss the fourfold symmetric vortex core structure 
by considering the effect of these correction terms. 

The purposes of this paper are, first, to construct the GL theory from the 
Gor'kov equation for the pure $d_{x^2-y^2}$-wave superconductors with 
including correction terms of the order $\ln(T_c/T)$ and, second, to 
investigate the fourfold symmetric core structure (i.e., pair potential, 
current and magnetic field) of an isolated single vortex by using this 
extended GL theory. 
We clarify how the vortex core structure deviates from the cylindrical 
symmetry. 
Here we consider the case of an isolated single vortex under a magnetic field 
applied parallel to the $c$-axis (or $z$-axis) in the clean limit. 
The Fermi surface is assumed to be two-dimensional, which is appropriate 
to high-$T_c$ superconductors, and isotropic in order to clarify effects 
of the $d$-wave nature of the pair potential on the vortex core structure. 
The additional anisotropy coming from, e.g., the Fermi surface can be 
incorporated into our extended GL framework.

The rest of this paper is organized as follows. 
In \S \ref{sec:2}, we construct the GL theory with including correction 
terms for the $d_{x^2-y^2}$-wave pairing. 
By using this GL theory, we investigate the fourfold symmetric vortex 
core structure. 
The pair potential is studied in \S \ref{sec:3}, and current and 
magnetic field in \S \ref{sec:4}. 
The summary and discussions are given in \S \ref{sec:5}. 
We set $\hbar =c =k_{\rm B} =1$ throughout the paper.

\section{GL theory for $d_{x^2-y^2}$-wave pairing}
\label{sec:2}

We consider the GL theory in pure $d_{x^2-y^2}$-wave superconductors.
The pair potential is given by 
\begin{equation}
\Delta({\mib r},{\mib k})= \eta({\mib r})\sqrt{2}\cos 2 \theta .
\label{eq:2.1}
\end{equation}
Here, ${\mib r}=(x,y)=(r \cos\phi, r \sin\phi )$ is the center of mass 
coordinate and ${\mib k}$ the relative coordinate of the Cooper pairs. 
Now, ${\mib k}$ is denoted by an angle $\theta$ measured from the $a$-axis 
(or $x$-axis) in the $ab$-plane. 
To study the effect of correction terms in the order of small parameter 
$\ln(T_c/T) \simeq 1- T/T_c$, we consider the GL theory by including the 
next higher order terms of $\eta$ and its derivatives other than those 
in the conventional GL theory.
The GL equation and the current density ${\mib j}$ are derived as follows 
from the Gor'kov equation within the weak coupling approximation (see the 
detailed derivation in Appendix A), 

\begin{eqnarray}
\ln \left( \frac{T_c}{T} \right) \eta 
&-& 
\beta \left( \eta|\eta|^2 - \frac{v_F^2}{6} {\mib D}^2\eta \right) 
+\alpha\beta^2 \biggl[ \frac{5}{6}\eta|\eta|^4  
+ \frac{v_F^4}{288} 
\left( 7D^4_x+\overline{D^2_xD^2_y}+7 D^4_y \right)\eta   \nonumber\\
&&
- \frac{v_F^2}{12}\left\{
4 |\eta|^2 ({\mib D}^2\eta) + \eta^2 ({\mib D}^2\eta)^\ast 
+2 \eta |{\mib D}\eta|^2 + 3({\mib D}\eta)^2\eta^\ast \right\} 
\biggr] +({\rm higher \ order \ terms})=0, 
\label{eq:2.2}
\end{eqnarray}
\begin{subequations}
\begin{eqnarray}
j_x= 
2 j_0 {\rm Im}  \biggl[(D_x \eta)\eta^\ast 
&&+\frac{v_F^2}{48}\alpha\beta 
\left\{ 7(D_x^3 \eta)\eta^\ast -7 (D_x^2 \eta)(D_x \eta)^\ast 
+(D_x \eta)(D_y^2 \eta)^\ast + (\overline{D_x D_y^2}\eta)\eta^\ast 
+(D_y \eta)(\overline{D_x D_y}\eta)^\ast \right\}  \nonumber\\
&& 
-3\alpha\beta|\eta|^2(D_x \eta) \eta^\ast 
+({\rm higher \ order \ terms})\biggr] , 
\label{eq:2.3a}
\end{eqnarray}
\begin{eqnarray}
j_y=
2 j_0 {\rm Im}  \biggl[(D_y \eta)\eta^\ast
&&+\frac{v_F^2}{48}\alpha\beta
\left\{ 7(D_y^3 \eta)\eta^\ast -7 (D_y^2 \eta)(D_y \eta)^\ast
+(D_y \eta)(D_x^2 \eta)^\ast + (\overline{D_x^2 D_y}\eta)\eta^\ast
+(D_x \eta)(\overline{D_x D_y}\eta)^\ast \right\}  \nonumber\\
&&
-3\alpha\beta|\eta|^2(D_y \eta) \eta^\ast 
+({\rm higher \ order \ terms}) \biggr], 
\label{eq:2.3b}
\end{eqnarray}
\label{eq:2.3}
\end{subequations}

\noindent 
where 
\begin{equation}
\alpha=\frac{62\zeta(5)}{49\zeta(3)^2}=0.908..., \qquad
\beta= \frac{21\zeta(3)}{16\pi^2T^2},  \qquad
j_0=-\frac{1}{3} |e| N_F v_F^2 \beta 
\label{eq:2.4}
\end{equation}
with Riemann's $\zeta$-functions $\zeta(3)$ and $\zeta(5)$, 
the Fermi velocity $v_F$ and the density of states at the Fermi surface $N_F$.
The differential operator ${\mib D}$ (${\mib D}^\ast$) is defined by 
${\mib D}={\mib \nabla}+i(2\pi / \phi_0){\mib A}$ 
(${\mib D}^\ast={\mib \nabla}-i(2\pi / \phi_0){\mib A}$)
with the vector potential ${\mib A}$ and the flux quantum $\phi_0$.
In Eqs. (\ref{eq:2.2}) and (\ref{eq:2.3}), $\overline{D_x^m D_y^n}$ means 
to take all the permutations of the product, such as $\overline{D_x^2 D_y}
=D_x^2 D_y + D_x D_y D_x + D_y D_x^2 $.

It is convenient to discuss Eqs.(\ref{eq:2.2}) and (\ref{eq:2.3}) 
in the following dimensionless unit, 
\begin{equation} 
\eta/\eta_0 \rightarrow \eta, \qquad r/\xi \rightarrow r, \qquad  
{\mib j}/(\eta_0^2/\xi)j_0 \rightarrow {\mib j}
\label{eq:2.5}
\end{equation}
with the GL coherent length $\xi=\{\beta v_F^2 /6 \ln(T_c/T)\}^{1/2}$  and 
the energy gap $\eta_0=\{\ln(T_c/T)/\beta \}^{1/2}$. 

In the dimensionless unit, the GL equation in Eq.(\ref{eq:2.2}) and ${\mib j}$
in Eq.(\ref{eq:2.3}), respectively, are written as 

\begin{eqnarray}
\eta -\eta|\eta|^2 +
&& 
{\mib D}^2\eta +\alpha\ln\left(\frac{T_c}{T}\right) 
\biggl[ \frac{5}{6}\eta|\eta|^4  
+\frac{1}{8} \left( 7D^4_x+\overline{D^2_xD^2_y}+7 D^4_y \right)\eta   
\nonumber\\
&&
- \frac{1}{2}\Bigl\{
4 |\eta|^2 ({\mib D}^2\eta) + \eta^2 ({\mib D}^2\eta)^\ast 
+ 2 \eta |{\mib D}\eta|^2 + 3 ({\mib D}\eta)^2\eta^\ast \Bigr\} 
\biggr] +O(\{\ln(T_c/T)\}^2 ) =0, 
\label{eq:2.6}
\end{eqnarray}
\begin{subequations}
\begin{eqnarray}
j_x
&=& 2 {\rm Im} 
\biggl[ \left\{ 1-3|\eta|^2\alpha\ln\left(\frac{T_c}{T}\right) \right\} 
(D_x \eta)\eta^\ast 
+ \frac{\alpha}{8}\ln\left(\frac{T_c}{T}\right) 
\Bigl\{ 7(D_x^3 \eta)\eta^\ast -7 (D_x^2 \eta)(D_x \eta)^\ast 
\nonumber \\
&& 
+(D_x \eta)(D_y^2 \eta)^\ast + (\overline{D_x D_y^2}\eta)\eta^\ast 
+(D_y \eta)(\overline{D_x D_y}\eta)^\ast \Bigr\} 
+ O(\{\ln(T_c/T)\}^2 ) \biggr]  , 
\label{eq:2.7a}
\end{eqnarray}
\begin{eqnarray}
j_y
&=& 2 {\rm Im} 
\biggl[ \left\{ 1-3|\eta|^2\alpha\ln\left(\frac{T_c}{T}\right) \right\}
(D_y \eta)\eta^\ast
+\frac{\alpha}{8}\ln\left(\frac{T_c}{T}\right)
\Bigl\{ 7(D_y^3 \eta)\eta^\ast -7 (D_y^2 \eta)(D_y \eta)^\ast
\nonumber \\
&&
+(D_y \eta)(D_x^2 \eta)^\ast + (\overline{D_x^2 D_y}\eta)\eta^\ast
+(D_x \eta)(\overline{D_x D_y}\eta)^\ast \Bigr\} 
+ O(\{\ln(T_c/T)\}^2 ) \biggr].
\label{eq:2.7b}
\end{eqnarray}
\label{eq:2.7}
\end{subequations}

\noindent 
In Eqs. (\ref{eq:2.6}) and (\ref{eq:2.7}), the neglected terms, 
which are higher 
order of $\eta$ and derivative, are in the order $\{\ln(T_c/T)\}^n$ $(n\ge 2)$.
We note that, in the limit $T \rightarrow T_c$, Eqs. (\ref{eq:2.6}) and 
(\ref{eq:2.7}) reduce to those of the conventional GL theory, and are  
the same forms as those of the conventional $s$-wave case. 
This indicates that there are no difference between $s$-wave and $d$-wave 
pairing within the conventional GL framework. 
Therefore, the vortex structure for $d$-wave pairing remains cylindrically  
symmetric within the conventional GL framework.  
The difference between $s$-wave and $d$-wave pairing first appears in the 
correction terms of the order $\ln(T_c/T)$.  
In particular, the $D^4$-terms in the GL equation (\ref{eq:2.6}) and 
the $D^3$-terms in the current density (\ref{eq:2.7}) are seen to break 
the cylindrical symmetry, and lead to the fourfold symmetry. 
These correction terms play an important role as the temperature decreases 
below the transition temperature.

\section{Pair potential around a single vortex}
\label{sec:3}
\subsection{Symmetry consideration}

The $d_{x^2-y^2}$-wave pairing, i.e., 
$\hat k_x^2 - \hat k_y^2 = \cos 2 \theta$ in momentum space, has fourfold 
symmetry for the rotation about the $c$-axis. 
Reflecting this property, the symmetry of a single vortex in 
$d_{x^2-y^2}$-wave superconductors is in the class $D^{(1)}(D_2)\times R$. 
In the limit $T \rightarrow T_c$, the GL equation (\ref{eq:2.6}) reduces to 
the conventional GL equation, which is the same as that of the isotropic  
$s$-wave pairing case. 
Therefore, following the well known consideration about a single 
vortex,~\cite{Fetter}  we obtain the structure of a single vortex as 
follows in the limit $T \rightarrow T_c$, 
\begin{equation}
\Delta({\mib r},{\mib k})= \eta({\mib r})\sqrt{2}\cos 2 \theta 
=f_0(r)e^{i\phi}\sqrt{2}\cos 2 \theta, 
\label{eq:3.1}
\end{equation}
which is the isotropic state with the winding number 1 around the vortex. 
As noted by Volovik,~\cite{Volovik}  
eight elements of the symmetry operations of the function 
$\Delta({\mib r},{\mib k})$ in Eq. (\ref{eq:3.1}) form the group $D_4(E)$: 
\begin{equation}
D_4(E)=\{ E,C_2e^{i\pi},U_{2,x}R,U_{2,y}e^{i\pi}R, C_4e^{i\pi/2}, 
C^3_4e^{-i\pi/2}, U_{2,x+y}e^{i\pi/2}R,  U_{2,x-y}e^{-i\pi/2}R\}. 
\label{eq:3.2}
\end{equation}
Here, $R$ is the time-reversal, $C_n$ the $(2 \pi/n)$-rotation about 
$z$-axis, and $U_{2,x}$, $U_{2,y}$, $U_{2,x+y}$ and $U_{2,x-y}$ denote 
the $\pi$-rotation about the $x$-axis, the $y$-axis, the lines $x+y=0$ 
and $x-y=0$, respectively.

On lowering temperature, other components with different winding numbers 
may be induced in $\eta({\mib r})$. 
If the symmetry $D_4(E)$ is conserved, the possible pair potential is 
generally restricted to the following form (brief derivation is given 
in Appendix B),  
\begin{equation}
\eta({\bf {\it r}})
=\sum_{n=-\infty}^{\infty}f_n(r)e^{i(4n+1)\phi}, 
\label{eq:3.3}
\end{equation}
where $f_n(r)$ is real. 
In the following, we determine the amplitude $f_n(r)$ 
from the GL equation (\ref{eq:2.6}).

\subsection{Pair potential around a single vortex}

We determine the amplitude $f_n(r)$ in Eq.(\ref{eq:3.3}) up to 
the order $\ln(T_c/T)$. 
Therefore we expand $f_n(r)$ in the powers of $\ln(T_c/T)$ as follows, 
\begin{equation}
f_n(r)=f_n^{(0)}(r)+ \alpha\ln(T_c/T) f_n^{(1)}(r)+O(\{\ln(T_c/T)\}^2 ), 
\label{eq:3.4}
\end{equation}
where $f_n^{(0)}(r)=0$ for $n \ne 0$ since $\Delta({\mib r},{\mib k})$ 
reduces to Eq. (\ref{eq:3.1}) in the limit $T \rightarrow T_c$.

We notice that, as we consider the isolated single vortex in the extreme 
type II superconductors (GL parameter $\kappa \gg 1$), the vector potential 
in ${\mib D}$ can be neglected.~\cite{dGennes}

Substituting Eq. (\ref{eq:3.4}) into Eq. (\ref{eq:2.6}) and writing 
the differential operators by the cylindrical coordinate, we obtain  
the following simultaneous differential equations, 
\begin{equation}
F^{(0)}_0(r) \equiv  
f^{(0)}_0(r)+\nabla^2(1)f^{(0)}_0(r)-f^{(0)}_0(r)^3=0 , 
\label{eq:3.5}
\end{equation}

\begin{eqnarray}
F^{(1)}_0(r) \equiv 
&&
f^{(1)}_0(r)+\nabla^2(1)f^{(1)}_0(r)-3f^{(0)}_0(r)^2f^{(1)}_0(r)
+\frac{3}{4} D_1(1)f^{(0)}_0(r) \nonumber \\
&&
-\frac{1}{2} \left( 5f^{(0)}_0(r)^2\nabla^2(1)f^{(0)}_0(r)
+5f^{(0)}_0(r){f^{(0)}_0}'(r)^2-\frac{1}{r^2}f^{(0)}_0(r)^3 \right)
+\frac{5}{6}f^{(0)}_0(r)^5=0, 
\label{eq:3.6}
\end{eqnarray}
\begin{equation}
F^{(1)}_1(r) \equiv
f^{(1)}_1(r)+\nabla^2(5)f^{(1)}_1(r)- \left( 2f^{(1)}_1(r)
+f^{(1)}_{-1}(r) \right) f^{(0)}_0(r)^2+\frac{1}{16} D_{+}(1)f^{(0)}_0(r)=0, 
\label{eq:3.7}
\end{equation}
\begin{equation}
F^{(1)}_{-1}(r) \equiv
f^{(1)}_{-1}(r)+\nabla^2(-3)f^{(1)}_{-1}(r)- \left( 2f^{(1)}_{-1}(r)
+f^{(1)}_{1}(r) \right) f^{(0)}_0(r)^2+\frac{1}{16} D_{-}(1)f^{(0)}_0(r)=0 , 
\label{eq:3.8}
\end{equation}
\begin{equation}
F^{(1)}_{n}(r) \equiv
f^{(1)}_{n}(r)+\nabla^2(4n+1)f^{(1)}_{n}(r)- \left( 2f^{(1)}_{n}(r)
+f^{(1)}_{-n}(r) \right) f^{(0)}_0(r)^2=0, \qquad (n\neq0,\pm 1)
\label{eq:3.9}
\end{equation}
where the differential  operators $\nabla^2(n)$, $D_1(n)$ and 
$D_{\pm}(n)$ are defined as 
\begin{equation}
\nabla^2(n) \equiv 
\frac{d^2}{dr^2}+\frac{1}{r} \frac{d}{dr} -\frac{n^2}{r^2} , 
\label{eq:3.10}
\end{equation}
\begin{equation}
D_1(n) \equiv 
\frac{d^4}{dr^4} +\frac{2}{r}\frac{d^3}{dr^3} 
-\frac{1}{r^2}\frac{d^2}{dr^2}+\frac{1}{r^3}\frac{d}{dr} 
-n^2 \left( \frac{2}{r^2}\frac{d^2}{dr^2} -\frac{2}{r^3}\frac{d}{dr} 
+\frac{4}{r^4} \right) + \frac{n^4}{r^4}, 
\label{eq:3.11}
\end{equation}
\begin{eqnarray}
D_{\pm}(n) \equiv 
&&
\frac{d^4}{dr^4} -\frac{6}{r}\frac{d^3}{dr^3} 
+\frac{15}{r^2}\frac{d^2}{dr^2} -\frac{15}{r^3}\frac{d}{dr} 
\mp 4n \left( \frac{1}{r}\frac{d^3}{dr^3} -\frac{6}{r^2}\frac{d^2}{dr^2} 
+\frac{14}{r^3}\frac{d}{dr} -\frac{12}{r^4} \right)  \nonumber \\
&&
+n^2 \left( \frac{6}{r^2}\frac{d^2}{dr^2} -\frac{30}{r^3}\frac{d}{dr} 
+\frac{44}{r^4} \right) 
\mp 4n^3 \left( \frac{1}{r^3}\frac{d}{dr} -\frac{3}{r^4} \right) 
+\frac{n^4}{r^4}.
\label{eq:3.12}
\end{eqnarray}

As for the boundary condition, 
\begin{equation}
f^{(0)}_0(0)=f^{(1)}_n(0)=0
\label{eq:3.13}
\end{equation}
at the vortex center so that divergence does not occur at 
$r \rightarrow 0$, and 
\begin{equation}
f^{(0)}_0(r \rightarrow \infty)=1, \qquad 
f^{(1)}_0(r \rightarrow \infty)=\frac{5}{12}, \qquad
f^{(1)}_n(r \rightarrow \infty)=0  \quad  (n\neq0) 
\label{eq:3.14}
\end{equation}
far from the vortex core, where the $r$-dependence of $\eta({\mib r})$ can be 
neglected ensuring $\eta({\mib r})$ to become the bulk value. 
Equation (\ref{eq:3.14}) is obtained by setting the terms with the 
differential operators to be 0 in Eqs. (\ref{eq:3.5}) - (\ref{eq:3.9}).
Equation (\ref{eq:3.5}) corresponds to the conventional GL equation, 
leading to the well known vortex core structure of the conventional 
GL theory for  $f^{(0)}_0(r)$.~\cite{Fetter}  
First, we obtain $f^{(0)}_0(r)$ from Eq. (\ref{eq:3.5}), and substitute it to 
Eqs. (\ref{eq:3.6}) - (\ref{eq:3.8}). 
Then, we obtain the correction terms $f^{(1)}_0(r)$ from Eq. (\ref{eq:3.6}), 
and $f^{(1)}_{\pm 1}(r)$ from Eqs. (\ref{eq:3.7}) and (\ref{eq:3.8}). 
From Eq. (\ref{eq:3.9}), we obtain $f^{(1)}_{n}(r)=0$ for $n \ne 0, \pm 1$. 
In the order $\ln(T_c/T)$, therefore, the components with $e^{-3i\phi}$ and 
$e^{5i\phi}$ are induced in addition to the isotropic components of 
$e^{i\phi}$. 
Components with other winding numbers are induced in the higher order of 
$\ln(T_c/T)$, for example, components with $e^{-7i\phi}$ and $e^{9i\phi}$ 
first appear in the order $\{ \ln(T_c/T) \}^2$.

The concrete form of the solution $f^{(j)}_i(r)$ in Eqs. (\ref{eq:3.5}) - 
(\ref{eq:3.9}) cannot obtained by the $r$- or $r^{-1}$-expansion. 
When we set 
\begin{equation}
f^{(j)}_i(r)=\sum_{m=1}^\infty C^{(j)}_{i,m} r^m 
\label{eq:3.15}
\end{equation}
to study the solution for $r \ll 1$, we obtain the following results 
for $m \le 3$, 
\begin{equation}
C^{(0)}_{0,m}=C^{(1)}_{0,m}=0, \quad (m \ne 1,3) \qquad
C^{(1)}_{-1,m}=0, \quad (m \ne 3)  \qquad
C^{(1)}_{i,m}=0 \ {\rm for} \ i \ne 0, -1, 
\label{eq:3.16}
\end{equation}
which means that $f_0(r)=O(r)$, $f_{-1}(r)=O(r^3)$ and $f_1(r)=O(r^4)$. 
However, the other coefficients $C^{(j)}_{i,m}$ cannot be determined uniquely 
in this $r$-expansion. 
On the other hand, when we set 
\begin{equation}
f^{(j)}_i(r)=\sum_{m=0}^\infty C^{(j)}_{i,-m} r^{-m}
\label{eq:3.17}
\end{equation}
to study the solution for $r \gg 1$, inconsistency among Eqs. 
(\ref{eq:3.5}) - (\ref{eq:3.9}) occurs in determining $C^{(1)}_{1,-4}$ 
and $C^{(1)}_{-1,-4}$. 
It suggests that the singular terms such as $r^{-1/2}e^{-r}$ probably exist in 
addition to $r^{-m}$-terms.

\subsection{Numerical Results}

We solve numerically the differential equations (\ref{eq:3.5}) - 
(\ref{eq:3.9}) by using the so-called relaxation method, which is often used 
in determining the vortex structure from the GL theory. 
The relaxation step 
\begin{equation}
f^{(j)}_i(r)^{[{\rm new}]}= f^{(j)}_i(r)^{[{\rm old}]} + c F^{(j)}_i(r) \qquad 
(c:{\rm constant})
\label{eq:3.18}
\end{equation}
is repeated until the condition $|F^{(j)}_i(r)|\ll 1$ is fulfilled for 
each $r$, where $F^{(j)}_i(r)$ is defined in Eqs. (\ref{eq:3.5}) -  
(\ref{eq:3.9}).

Figure \ref{fig:1} (a)  shows $f^{(1)}_0(r)$, which is the correction to the 
factor of the isotropic component $e^{i\phi}$. 
As seen from Eq. (\ref{eq:3.4}), the correction 
$\alpha\ln(T_c/T)f^{(1)}_0(r)$ is added to the solution of 
the conventional GL theory $f^{(0)}_0(r)$.
In Fig. \ref{fig:1} (b), $f^{(1)}_1(r)$ and $f^{(1)}_{-1}(r)$ are presented, 
where $f^{(1)}_{-1}(r)$ behaves as $r^3$ near the core while 
$f^{(1)}_{1}(r)$ as $r^4$. 
These components induce the fourfold symmetric structure of the pair 
potential. 
They both attain a maximum at several coherence length from the origin,  
and smoothly and slowly fall off. 
It is to be noted that they never  drop exponentially. 
It indicates that the effect of the fourfold symmetric structure becomes small 
but remains far from the vortex core, and may affect a vortex-vortex 
interaction and the formation of a stable vortex lattice. 
This point may become important when considering a stable vortex lattice 
and its orientation relative to the underlying crystal in $d$-wave 
superconductors. 
It is confirmed that the other components  $f^{(1)}_n(r)$ ($|n|\ge2$) 
reduce to 0 in the relaxation method.

\end{full}
\begin{figure}[t]
\begin{full}
\begin{center}
\epsfig{file=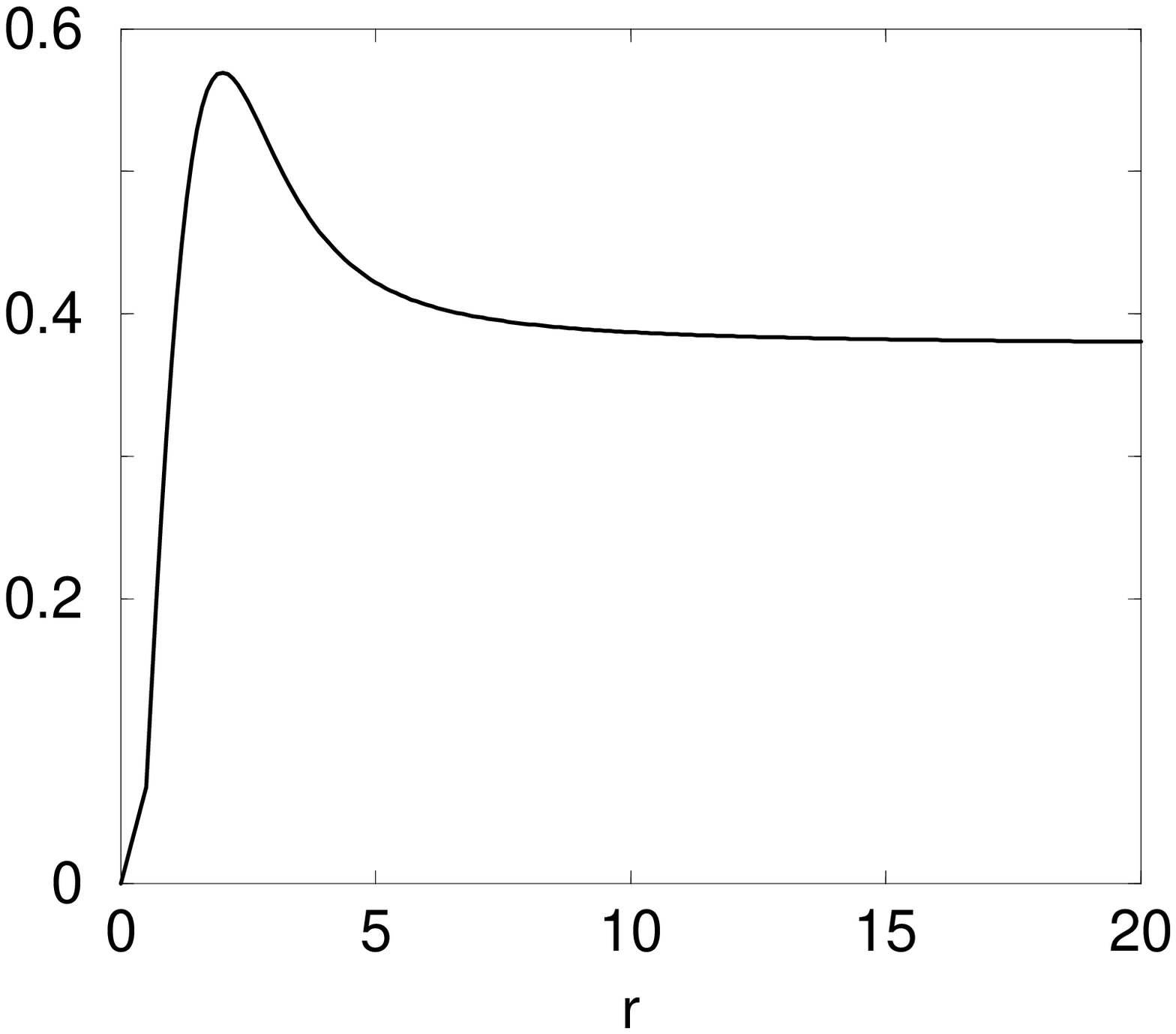, width=8.0cm}
\epsfig{file=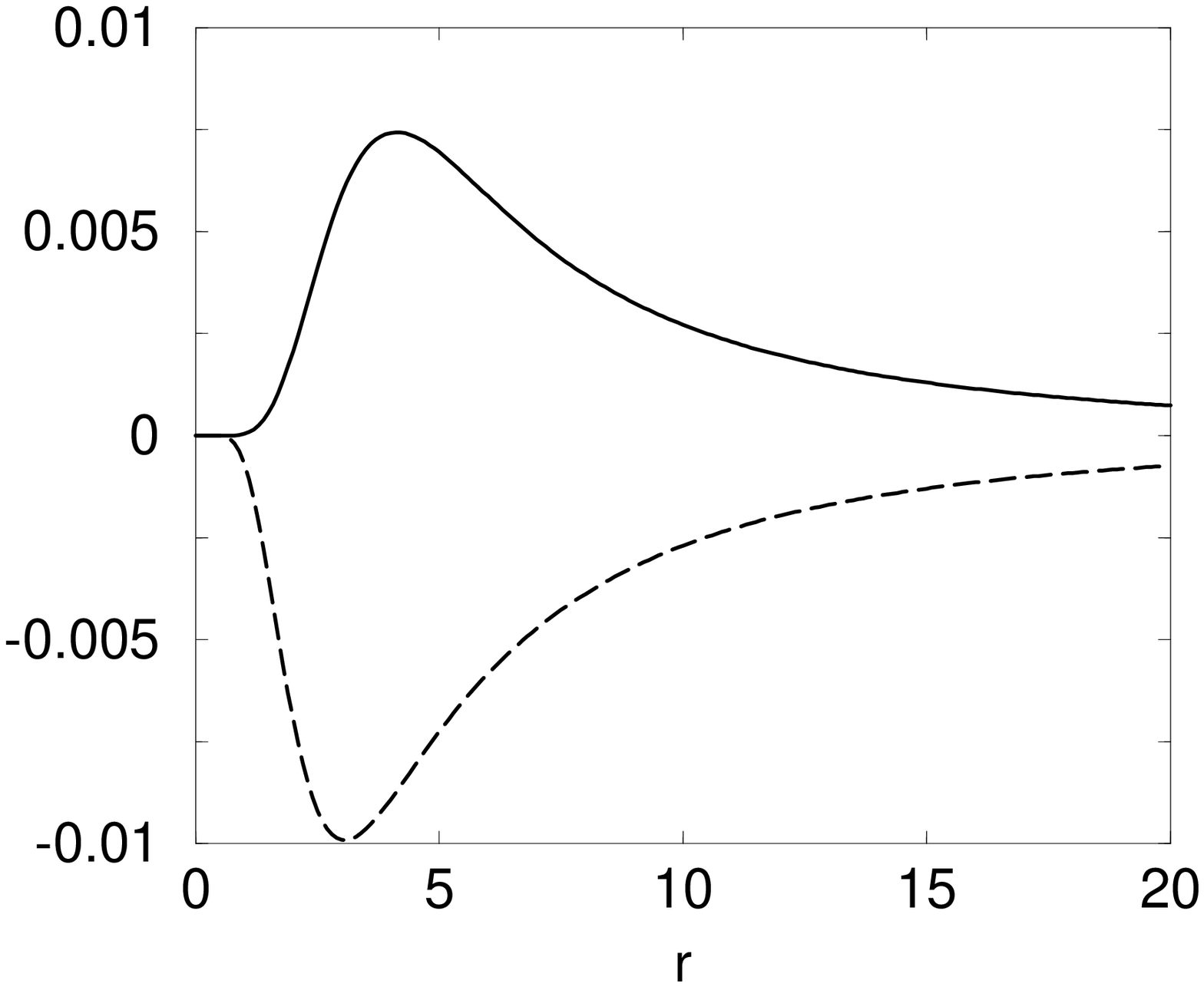, width=8.0cm}

(a)\hspace{7cm}(b)
\end{center}
\caption{
The spatial variation of the components of the order parameter 
around a vortex in the dimensionless unit:  
(a) $f^{(1)}_0(r)$, (b) $f^{(1)}_1(r)$ (solid curve) and 
$f^{(1)}_{-1}(r)$ (long dashed curve), associated with 
the phase factor $e^{i \phi}$, $e^{5i \phi}$ and $e^{-3i \phi}$, respectively. 
}
\label{fig:1}
\end{full}
\end{figure}
\begin{full}

Using these solutions, the pair potential around a single vortex is written as 

\begin{equation}
\eta({\mib r})=\left\{ f^{(0)}_0(r)+ \alpha\ln(T_c/T)\left( f^{(1)}_0(r) 
+ f^{(1)}_1(r)e^{4i\phi} +f^{(1)}_{-1}(r)e^{-4i\phi} \right)\right\}e^{i\phi} 
+O(\{ \ln(T_c/T)\}^2). 
\label{eq:3.19}
\end{equation}
Then, the amplitude of the pair potential around a single vortex is given by 
\begin{eqnarray}
|\eta({\mib r})|
&=&
\Bigl[ \left\{ f^{(0)}_0(r)+ \alpha\ln(T_c/T) \left\{ f^{(1)}_0(r)
+\cos 4 \phi \left( f^{(1)}_1(r) +f^{(1)}_{-1}(r) \right) \right\}  \right\}^2 
\nonumber \\
&&
+ \left\{ \alpha\ln(T_c/T) \sin 4 \phi 
\left( f^{(1)}_1(r) - f^{(1)}_{-1}(r) \right) \right\}^2 \Bigr]^{1/2} 
\nonumber \\
&=&
f^{(0)}_0(r)+ \alpha\ln(T_c/T) \left\{ f^{(1)}_0(r) 
+\cos 4 \phi \left( f^{(1)}_1(r) +f^{(1)}_{-1}(r) \right) \right\} 
+O(\{ \ln(T_c/T)\}^2).
\label{eq:3.20}
\end{eqnarray}

\noindent 
As for the amplitude $|\eta({\mib r})|$ in Eq. (\ref{eq:3.20}), we see that 
the deviation from the cylindrical structure behaves like $\cos 4\phi$ 
around a vortex in the order $\ln(T_c/T)$. 
In Fig. \ref{fig:2} (a), we show the $r$-dependence of the difference of the 
amplitude along the $0^\circ$ direction (along the $x$-axis and $y$-axis) 
and along the $45^\circ$ direction (along the line $y=\pm x$), i.e., 
\begin{equation}
|\eta(r,\phi=0)|-|\eta(r,\phi=\pi/4)| 
= 2\alpha\ln(T_c/T)\left( f^{(1)}_1(r) +f^{(1)}_{-1}(r) \right) .  
\label{eq:3.21}
\end{equation}
In the figure, the factor of $\alpha\ln(T_c/T)$, i.e., 
$2( \, f^{(1)}_1(r) +f^{(1)}_{-1}(r) \, )$ is plotted. 
It indicates that the anisotropy localizes around the core region, and 
that the amplitude along the $0^\circ$ direction is suppressed compared 
with that  along the $45^\circ$ direction.

\end{full}
\begin{figure}[t]
\begin{full}
\begin{center}
\epsfig{file=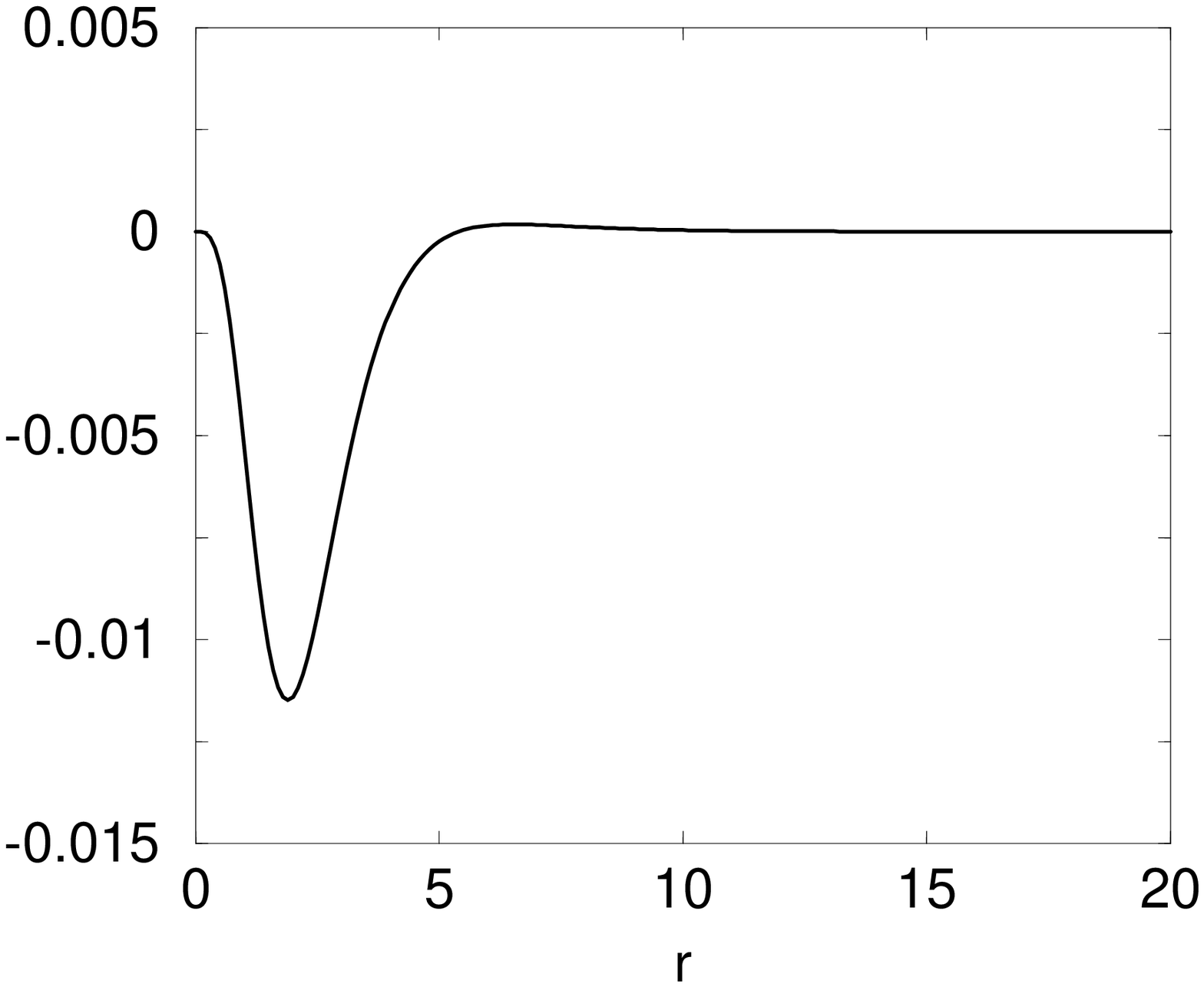, width=8.0cm}
\epsfig{file=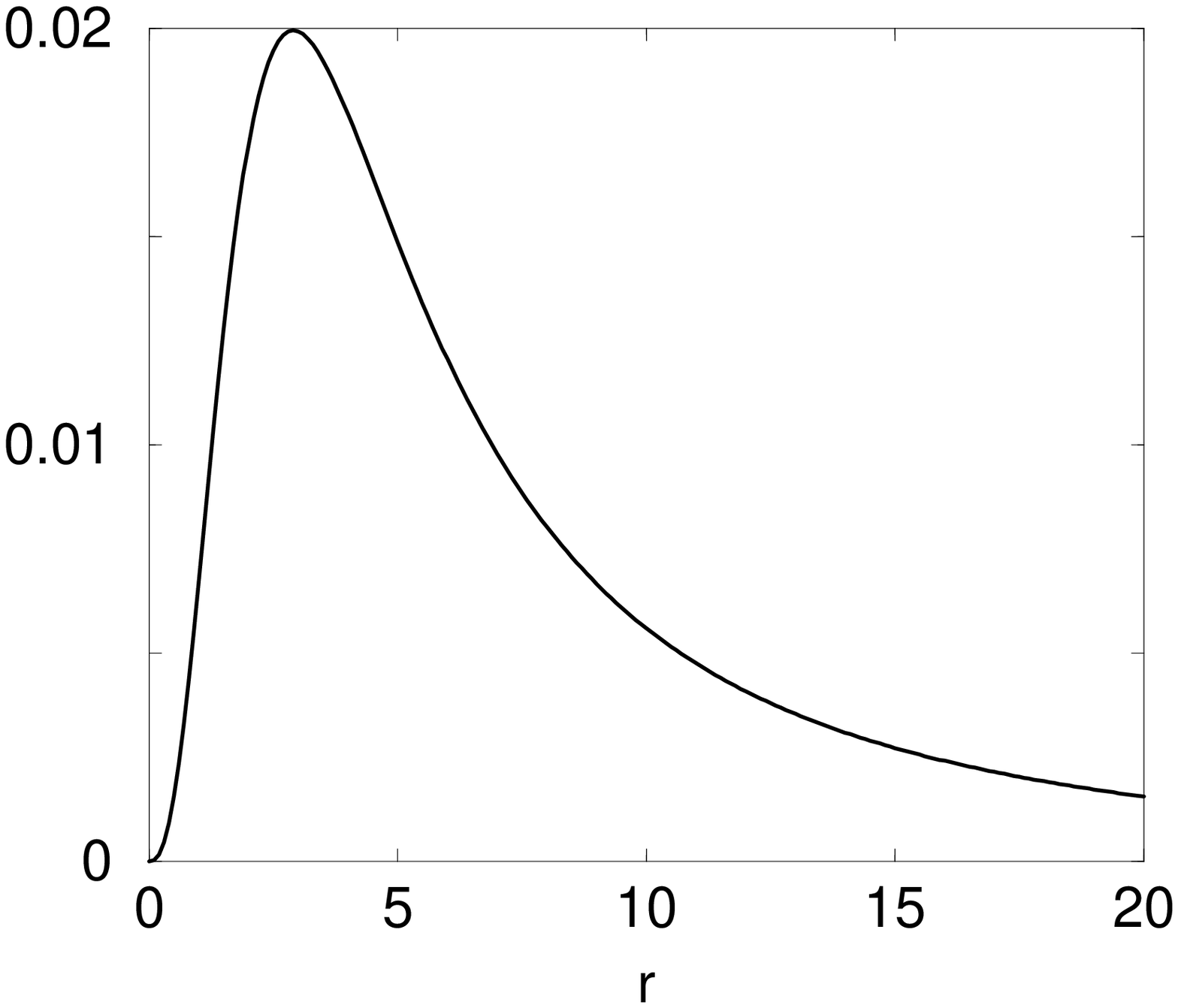, width=8.0cm}

(a)\hspace{7cm}(b)
\end{center}
\caption{
The spatial profile of the order parameter around a vortex.
(a) The deviation of the amplitude $|\eta(r,\phi)|$ at $\phi=0^\circ$ 
from $|\eta(r,\phi)|$ at $\phi=45^\circ$ along the radial direction,  
where the factor of $\alpha\ln(T_c/T)$ in Eq. (\ref{eq:3.21}) is displayed. 
(b) The correction of the phase, $\Phi(r,\phi)$,  at $\phi=22.5^\circ$, 
where the factor of $\alpha\ln(T_c/T)$ in Eq. (\ref{eq:3.23}) is displayed. 
}
\label{fig:2}
\end{full}
\end{figure}
\begin{full}

As for the phase of the pair potential, we consider the correction of 
the phase to the leading term which goes like $e^{i\phi}$. 
Then, we introduce $\Phi({\mib r})$ defined by
\begin{equation}
\eta({\mib r})=|\eta({\mib r})|e^{i\phi+i\Phi({\mib r})}.
\label{eq:3.22}
\end{equation}
From Eq.(\ref{eq:3.19}), the correction of the phase is given by

\begin{eqnarray}
\Phi({\mib r})
&=&
\tan^{-1}\left( 
\frac{\alpha\ln(T_c/T) \sin 4 \phi 
\left( f^{(1)}_1(r) - f^{(1)}_{-1}(r) \right)}
{ f^{(0)}_0(r)+ \alpha\ln(T_c/T) \left\{ f^{(1)}_0(r)
+\cos 4 \phi \left( f^{(1)}_1(r) +f^{(1)}_{-1}(r) \right) \right\} } \right)
\nonumber \\
&=&
\alpha\ln(T_c/T) \sin 4 \phi \left( f^{(1)}_1(r)-f^{(1)}_{-1}(r) \right)
\Big/ f^{(0)}_0(r) 
+O(\{ \ln(T_c/T)\}^2).
\label{eq:3.23}
\end{eqnarray}

\noindent 
It indicates that $\Phi({\mib r})$ behaves like $\sin 4\phi$ around a 
vortex in the order $\ln(T_c/T)$. 
In Fig. \ref{fig:2} (b), we show the $r$-dependence of $\Phi({\mib r})$ 
along the $22.5^\circ$ direction from the $a$ axis, 
i.e., $\Phi(r,\phi=\pi/8)$. 
In the figure, the factor of $\alpha\ln(T_c/T)$, i.e., 
$( \,  f^{(1)}_1(r)-f^{(1)}_{-1}(r) \, ) / f^{(0)}_0(r) $ is plotted. 
It is seen that the anisotropy of the phase rather extends compared to that 
of the amplitude in Fig. \ref{fig:2} (a) , 
and that $\Phi({\mib r})>0$ for the region $0 < \phi < \pi/4$.

The above-mentioned behavior of the pair potential $\eta({\mib r})$ 
qualitatively agrees well with that obtained in the quasi-classical 
Eilenberger theory (see Figs. 1 and 2 in Ref. \citen {IchiokaF} 
for comparison).

\section{Current and magnetic field around a vortex}
\label{sec:4}

The current density and the magnetic field around a single vortex is 
considered with the correction terms up to the order $\ln(T_c/T)$. 
Substituting Eq. (\ref{eq:3.19}) to Eq. (\ref{eq:2.7}), we obtained the 
radial component, $j_r = j_x \cos\phi + j_y \sin\phi$, and the rotational 
component, $j_\phi = -j_x \sin\phi + j_y \cos\phi$, of the current 
density as follows, 
\begin{equation}
j_r({\mib r})=
\alpha\ln(T_c/T)\sin 4\phi \Bigl( j_1(r) + j_2(r) \Bigr) 
+O(\{ \ln(T_c/T)\}^2) ,\label{eq:4.1}
\end{equation}
\begin{equation}
j_{\phi}({\mib r})=
\frac{2}{r}f^{(0)}_{0}(r)^2 
+\alpha\ln(T_c/T)\Bigl\{ -j_3(r) 
+\cos 4\phi \Bigl( j_1(r) +j_4(r) \Bigr)\Bigr\} 
+O(\{ \ln(T_c/T)\}^2), 
\label{eq:4.2}
\end{equation}
where 

\begin{equation}
j_1(r) \equiv 
-\frac{5}{2r^3}f^{(0)}_0(r)^2 +\frac{2}{r^2}{f^{(0)}_0}'(r)f^{(0)}_0(r) 
+\frac{1}{2r}\left( {f^{(0)}_0}'(r)^2 -2{f^{(0)}_0}''(r)f^{(0)}_0(r) \right),
\label{eq:4.3}
\end{equation}
\begin{equation}
j_2(r) \equiv 
2 \left( f^{(0)}_0(r){f^{(1)}_1}'(r)-f^{(0)}_0(r){f^{(1)}_{-1}}'(r)
+f^{(1)}_{-1}(r){f^{(0)}_{0}}'(r)-f^{(1)}_{1}(r){f^{(0)}_{0}}'(r) \right), 
\label{eq:4.4}
\end{equation}
\begin{equation}
j_3(r) \equiv
\frac{3}{r^3}f^{(0)}_{0}(r)^2 -\frac{4}{r^2}{f^{(0)}_{0}}'(r)f^{(0)}_{0}(r) 
+\frac{1}{r} \left( {f^{(0)}_{0}}'(r)^2 -2{f^{(0)}_{0}}''(r)f^{(0)}_{0}(r)
+3{f^{(0)}_{0}}'(r)^4 \right) 
-\frac{4}{r}{f^{(0)}_{0}}'(r){f^{(1)}_{0}}'(r), 
\label{eq:4.5}
\end{equation}
\begin{equation}
j_4(r) \equiv 
\frac{4}{r}f^{(0)}_0(r) \left( 3f^{(1)}_1(r)-f^{(1)}_{-1}(r) \right).
\label{eq:4.6}
\end{equation}
From Eqs. (\ref{eq:4.1}) and (\ref{eq:4.2}), the amplitude of the current 
density is given by
\begin{equation}
|{\mib j}({\mib r})|=
\frac{2}{r}f^{(0)}_{0}(r)^2 
+\alpha\ln(T_c/T)\Bigl\{ -j_3(r) 
+\cos 4\phi \Bigl( j_1(r) +j_4(r) \Bigr)\Bigr\} 
+O(\{ \ln(T_c/T)\}^2), 
\label{eq:4.7}
\end{equation}
which is the same expression as $j_\phi$ in Eq. (\ref{eq:4.2}) within 
the order $\ln(T_c/T)$. 
In the limit $T \rightarrow T_c$, Eqs. (\ref{eq:4.1}), (\ref{eq:4.2}) and 
(\ref{eq:4.7}) give the cylindrically symmetric structure of 
the conventional GL theory. 
This profile of $|{\mib j}({\mib r})|$ is shown in Fig. \ref{fig:3} (a) 
as a function of $r$.

From Eq. (\ref{eq:4.7}), we see that the deviation of $|{\mib j}({\mib r})|$ 
from the cylindrical structure first appears in the order $\ln(T_c/T)$ 
and behaves like $\cos 4\phi$ around a vortex.  
In Fig. \ref{fig:3} (b), we show the $r$-dependence of the difference of 
$|{\mib j}({\mib r})|$ along the $0^\circ$ direction and along the $45^\circ$ 
direction, i.e.,  
\begin{equation}
|{\mib j}(r,\phi=0)|-|{\mib j}(r,\phi=\pi/4)| 
= 2\alpha\ln(T_c/T)\Bigl( j_1(r) +j_4(r) \Bigr) . 
\label{eq:4.8}
\end{equation}
In the figure, the factor of $\alpha\ln(T_c/T)$, i.e., 
$2(\, j_1(r) +j_4(r) \, )$ is plotted. 
It is seen that $|{\mib j}(r,\phi=0)|<|{\mib j}(r,\phi=\pi/4)|$ near the 
vortex core and $|{\mib j}(r,\phi=0)|>|{\mib j}(r,\phi=\pi/4)|$ far from 
the core. 
It should be noted that this sign change in Fig. \ref{fig:3} (b) occurs 
at a point where $f^{(1)}_{-1}(r)$ in Fig. \ref{fig:1} (b) reaches the 
minimum. 

\end{full}
\begin{figure}[t]
\begin{full}
\begin{center}
\epsfig{file=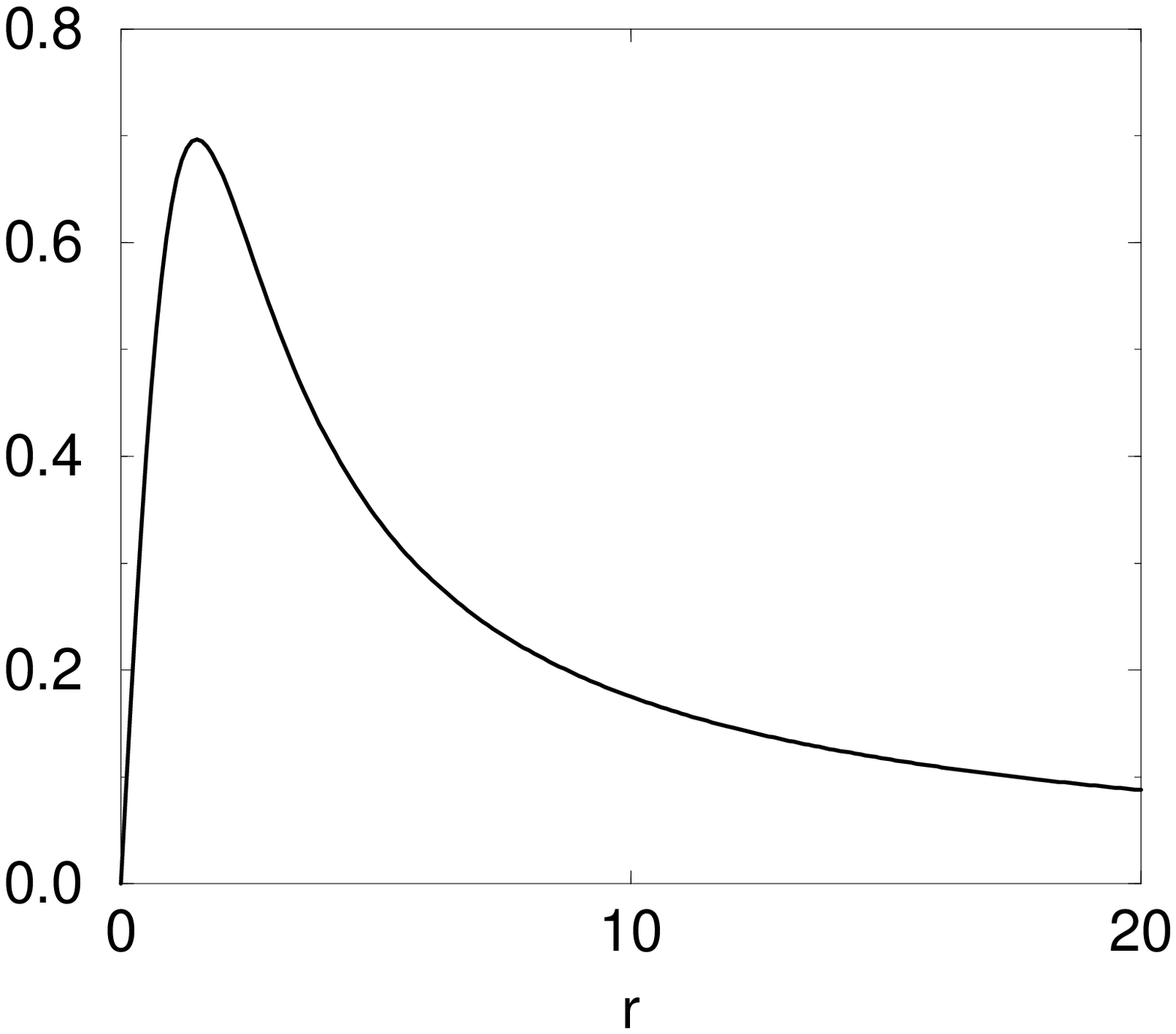, width=8.0cm}
\epsfig{file=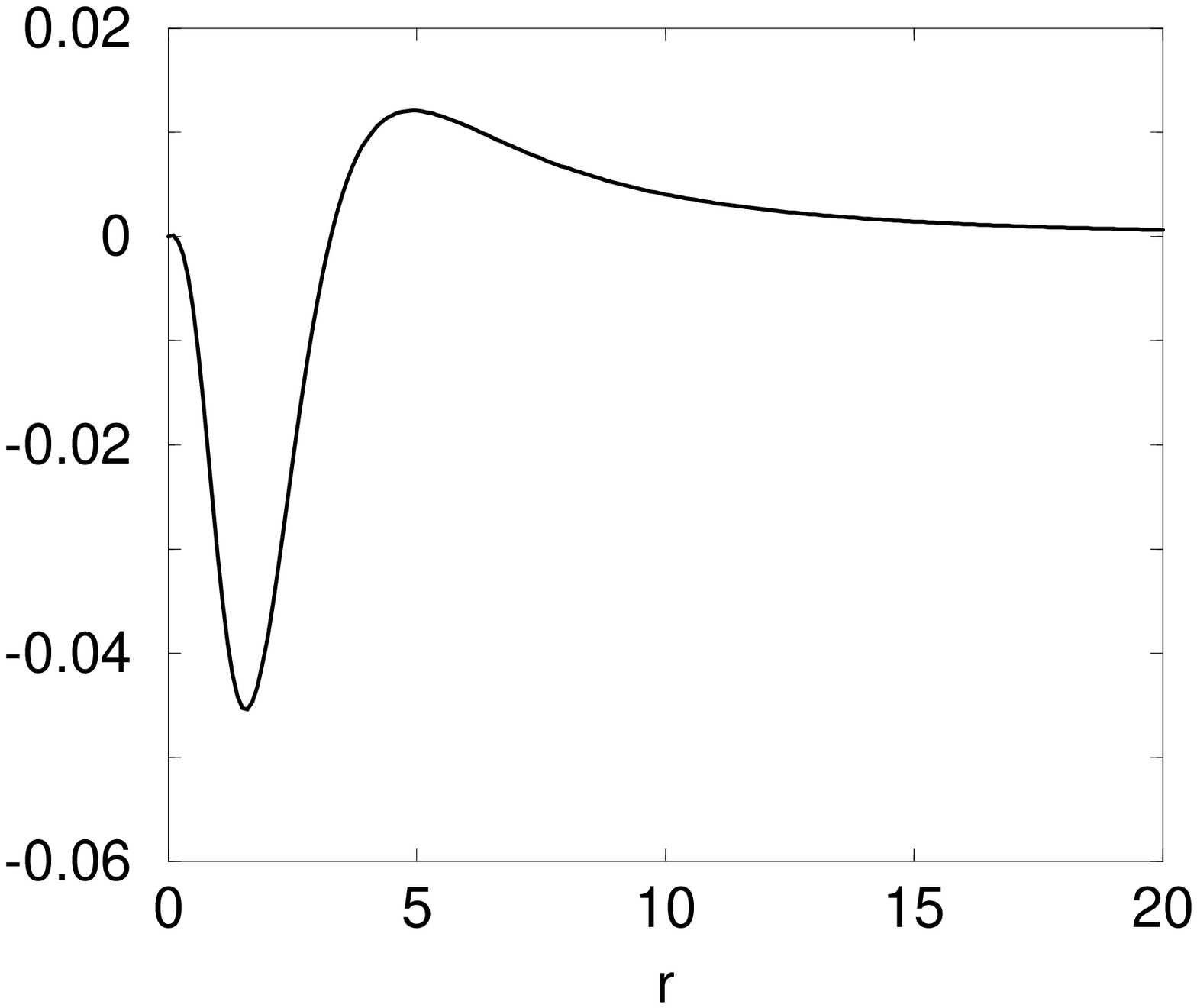, width=8.0cm}

(a)\hspace{7cm}(b)
\end{center}
\caption{
The spatial profile of the current density around a vortex. 
(a) The amplitude $|{\mib j}({\mib r})|$ as a function of $r$ 
in the limit $T \rightarrow T_c$.
(b) The deviation of $|{\mib j}(r,\phi)|$ at $\phi=0^\circ$ from 
$|{\mib j}(r,\phi)|$ at $\phi=45^\circ$ as a function of $r$, 
where the factor of $\alpha\ln(T_c/T)$ in Eq. (\ref{eq:4.8}) is displayed.  
This plot shows the deviation of $|{\mib j}({\mib r})|$ from the cylindrical 
structure. 
It is noted that the sign of the deviation changes at $r=3.2$ from negative 
to positive. 
}
\label{fig:3}
\end{full}
\end{figure}
\begin{full}

The resulting magnetic field ${\mib b}({\mib r})$ can be found from 
the Maxwell equation: 
\begin{equation}
{\mib j}({\mib r})= \kappa^2 {\mib \nabla}\times{\mib b}({\mib r})
\label{eq:4.9}
\end{equation}
in the dimensionless unit: Eq. (\ref{eq:2.5}) and 
\begin{equation}
{\mib b}({\mib r})/\sqrt{2} H_c \kappa \rightarrow {\mib b}({\mib r}), 
\label{eq:4.10}
\end{equation}
where $H_c=1/\kappa \xi^2 2 \sqrt{2}$ is the thermodynamic critical field. 
From Eqs. (\ref{eq:4.1}), (\ref{eq:4.2}) and (\ref{eq:4.9}), 
the magnetic field, which has only the $z$-component, is given by 
\begin{eqnarray}
b_z({\mib r})=
&&
b_z(r=0) -\frac{1}{\kappa^2}\int^r_0 \frac{2}{r'}f^{(0)}_0(r')^2 dr' 
+\frac{1}{4\kappa^2}\alpha\ln(T_c/T)\Bigl\{ 
(1-\cos 4\phi)r \Bigl( j_1(r)+j_2(r) \Bigr) \nonumber \\
&&
-4 \int^r_0 \Bigl( j_1(r') -j_3(r') +j_4(r') \Bigr) dr' \Bigl\} 
+O(\{ \ln(T_c/T)\}^2 ). 
\label{eq:4.11}
\end{eqnarray}
In the limit $T \rightarrow T_c$, Eq. (\ref{eq:4.11}) gives the 
cylindrically symmetric structure of the conventional GL theory. 
This profile of $b_z({\mib r})-b_z(r=0)$ is shown in Fig. \ref{fig:4} (a) 
as a function of $r$.

From Eq. (\ref{eq:4.11}), we find that the deviation of $b_z({\mib r})$ 
from the cylindrical structure first appears in the order $\ln(T_c/T)$ 
and behaves like $\cos 4\phi$ around a vortex.  
In Fig. \ref{fig:4} (b), we show the $r$-dependence of the difference of 
$b_z({\mib r})$ along the $0^\circ$ direction and along the $45^\circ$ 
direction, i.e., 
\begin{equation}
b_z(r,\phi=0)-b_z(r,\phi=\pi/4)
= -\frac{1}{2\kappa^2}\alpha\ln(T_c/T) r \Bigl( j_1(r)+j_2(r) \Bigr). 
\label{eq:4.12}
\end{equation}
In the figure, the factor of $\kappa^{-2} \alpha \ln(T_c/T)$, i.e., 
$- r ( \, j_1(r)+j_2(r)  \, )/2$ is plotted. 
It indicates that $b_z({\mib r})$ is large along the $0^\circ$ direction  
compared with along the $45^\circ$ direction, that is, the magnetic field 
extends along the $0^\circ$ direction around a vortex. 
It is noted that, reflecting the sign change in Fig. \ref{fig:3} (b), 
the anisotropy of $b_z({\mib r})$ becomes weaker far from the vortex core.

\end{full}
\begin{figure}[t]
\begin{full}
\begin{center}
\epsfig{file=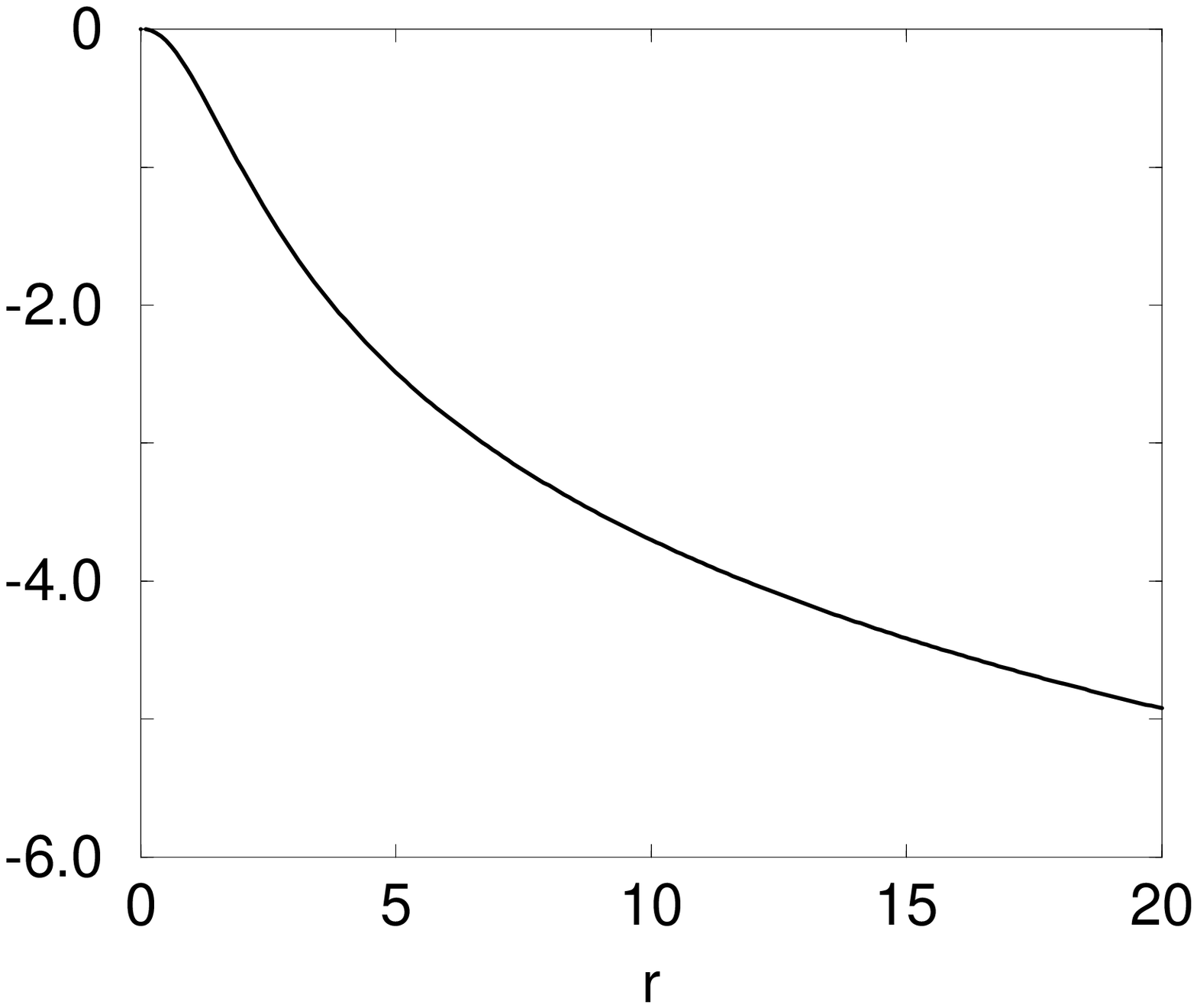, width=8.0cm}
\epsfig{file=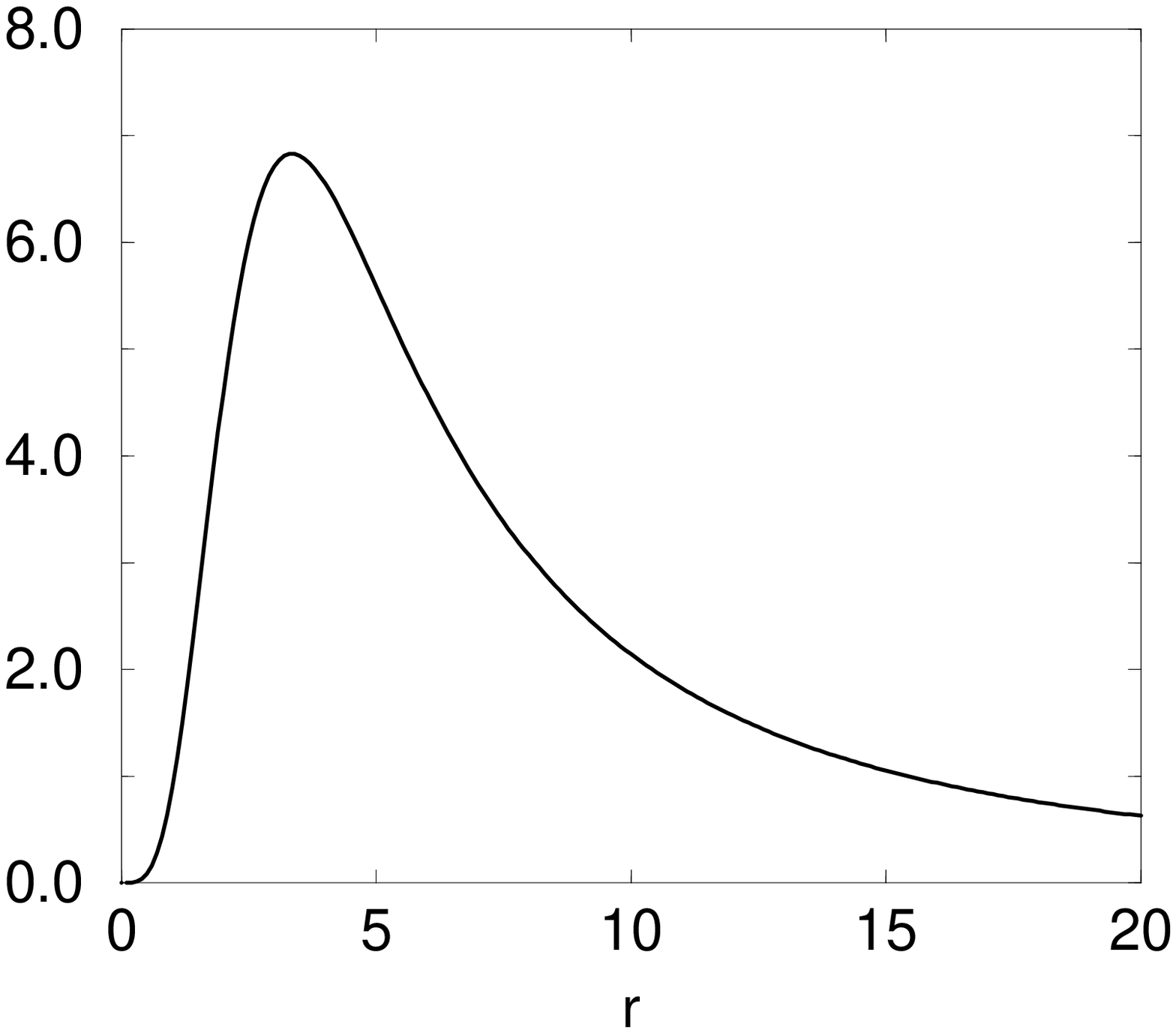, width=8.0cm}

(a)\hspace{7cm}(b)
\end{center}
\caption{
The spatial profile of the magnetic field around a vortex. 
(a) The magnetic field $b_z({\mib r})-b_z(r=0)$ as a function of $r$ in the 
limit $T \rightarrow T_c$, where the factor of $\kappa^{-2}$ in 
Eq. (\ref{eq:4.11}) is displayed. 
(b) The deviation of $b_z(r,\phi)$ at $\phi=0^\circ$ from
$b_z(r,\phi)$ at $\phi=45^\circ$ as a function of $r$,
where the factor of $\kappa^{-2} \alpha\ln(T_c/T)$ in Eq. (\ref{eq:4.12}) 
is displayed. 
This plot shows the deviation of $b_z({\mib r})$ from the cylindrical
structure.
Far from the vortex core, this deviation decreases and the fourfold 
symmetric structure becomes weaker.
}
\label{fig:4}
\end{full}
\end{figure}
\begin{full}

The above-mentioned behaviors of ${\mib j}({\mib r})$ and $b_z({\mib r})$ 
qualitatively agree well with those obtained in the quasi-classical
Eilenberger theory (see Figs. 4 - 7 in Ref. \citen {IchiokaF}
for comparison).

\section{Summary and discussions}
\label{sec:5}

First, the GL theory is constructed from the Gor'kov equation with including 
the correction terms up to the order $\ln(T_c/T)$ in pure 
$d_{x^2-y^2}$-wave superconductors. 
In the limit $T \rightarrow T_c$, the GL equation and the current density in   
the $d_{x^2-y^2}$-wave case reduce to the same form as that 
of the isotropic $s$-wave superconductors. 
Thus, within the conventional GL framework, the vortex core structure 
remains cylindrically symmetric even in the $d_{x^2-y^2}$-wave case. 
The difference between the $s$-wave and $d_{x^2-y^2}$-wave first appears 
in the correction terms of the order $\ln(T_c/T)$. 

Second, by using this extended GL theory, we investigate the fourfold 
symmetric vortex core structure in the $d_{x^2-y^2}$-wave superconductors. 
Among the correction terms, there are some terms 
which break the cylindrical symmetry and have fourfold symmetry, which 
is the fact reflecting the fourfold symmetry of the Cooper pair, 
$\hat k_x^2 - \hat k_y^2$. 
Then, the vortex core structure becomes fourfold symmetric. 
We study the fourfold symmetric structure of the pair potential, 
the current and the magnetic field around an isolated single vortex 
with the correction terms up to the order $\ln(T_c/T)$. 
As a result, the amplitude of the pair potential is suppressed along the 
$0^\circ$ direction compared with along the $45^\circ$ direction.  
The phase of the pair potential also shows fourfold symmetry around a vortex. 
Along the $45^\circ$ direction, the amplitude of the current is enhanced 
near the vortex core and suppressed far away from the vortex core compared 
with along the $0^\circ$ direction. 
Reflecting this behavior of the current, the magnetic field is enhanced along 
the $0^\circ$ direction near the vortex core, and this fourfold symmetry of 
the magnetic field fades out slowly far from the vortex core. 
These fourfold symmetric structures of the pair potential, the current and 
the magnetic field qualitatively agree well with the results by the 
quasi-classical Eilenberger theory.~\cite{IchiokaF} 
On lowering temperature, the correction terms of the order $\ln(T_c/T)$ 
make important roles, and the fourfold symmetric vortex core structure 
becomes clear. 

It should be noted that, while our obtained vortex core structure has similar 
fourfold symmetry to that predicted by Xu {\it et al.}~\cite{Xu} from the 
two-component GL equation, the origin of the fourfold symmetry is quite 
different from theirs. 
In their theory, the fourfold symmetry is induced by the mixing of the 
$s$-wave component. 
Then, in the case induced $s$-wave component is negligibly small, i.e., 
the pure $d_{x^2-y^2}$-wave case, the vortex core structure reduces to 
cylindrically symmetric one. 
On the other hand, we consider the pure $d_{x^2-y^2}$-wave case. 
Even in this case, the vortex core structure exhibits fourfold symmetry 
in our theory. 
It is due to the fourfold symmetric terms which are higher order of 
$\ln(T_c/T)$ and neglected in the conventional GL theory. 

In connection with these correction terms, 
Takanaka and Kuboya~\cite{Takanaka} and Won and Maki~\cite{Won,Maki} 
explained the fourfold symmetric behavior of $H_{c2}$ in the $ab$-pane 
and the deformation of the vortex lattice from a triangular lattice 
by considering the effect of these terms. 

By the extended GL theory with the correction terms, we succeed in 
reproducing qualitatively the fourfold symmetric vortex core structure 
obtained by the quasi-classical Eilenberger theory.~\cite{IchiokaF} 
This fact indicates that the correction terms of the order $\ln(T/T_c)$ 
are important in the GL framework, when we consider 
the vortex structure in pure $d_{x^2-y^2}$-wave superconductors. 
The GL framework presented here is found to be useful to analyze the 
vortex core structure and to obtain further physical insight. 
While enormous numerical computations are needed in the Eilenberger theory, 
the GL theory gives the information about the vortex structure with 
concise calculations.

As shown in Fig. \ref{fig:1} (b), the effect of the fourfold symmetric 
vortex core structure becomes small but remains far from the vortex core. 
Then it may affect a vortex-vortex interaction and the formation of a 
stable vortex lattice. 
This point may become important when considering a stable vortex lattice 
and its orientation relative to the underlying crystal in $d$-wave 
superconductors, which belongs to our future study.

\section*{Acknowledgments}
The authors thank K. Takanaka, B. Keimer, Ch.  Renner  
and \O. Fischer for useful discussions and information. 

\appendix
\section{Derivation of the GL equation}
\label{sec:ap.a}

The GL equation (\ref{eq:2.2}) and the current expression (\ref{eq:2.3}) 
in the $d_{x^2-y^2}$-wave pairing case 
are derived from the Gor'kov equation with including both the higher power 
of the pair potential and higher order derivative. 
Here, we consider the singlet anisotropic pairing case, in general. 
Similar derivation was presented by Takanaka and Nagashima in the case of 
isotropic $s$-wave pairing but anisotropic density of states at the Fermi 
surface.~\cite{TakanakaPTP}  
In our derivation, higher order terms are considered up to the order 
${\mib D}^m \eta^n$ $(m+n \le 5)$, and the integral $\int d{\mib r}$ and 
$\int d{\mib k}/(2 \pi)^2$ are noted by $\sum_{\mibs r}$ and $\sum_{\mibs k}$, 
respectively.

As for the anisotropic pairing, the pair potential has the non-local form 
$\Delta({\mib r},{\mib r}')$. 
Thus, in the Gor'kov equation for the Green function 
$G_{\omega_n}({\mib r},{\mib r}')$ and the anomalous Green functions 
$F_{\omega_n}({\mib r},{\mib r}')$ and  
$F^{\dagger}_{\omega_n}({\mib r},{\mib r}')$, 
the term $\Delta({\mib r})F_{\omega_n}({\mib r},{\mib r}')$ in 
the isotropic $s$-wave pairing case~\cite{AGD} is replaced by 
$\sum_{{\mibs r}''}\Delta({\mib r},{\mib r}'')
F_{\omega_n}({\mib r}'',{\mib r}')$, and the self-consistent condition is 
modified to 
\begin{equation}
\Delta^\ast ({\mib r},{\mib r}')
=V({\mib r}-{\mib r}')T\sum_{\omega_n}
F^{\dagger}_{\omega_n}({\mib r},{\mib r}') . 
\label{eq:a.1}
\end{equation}
From the Gor'kov  equation, we obtain the relations 
\begin{equation}
G_{\omega_n}({\mib r},{\mib r}') =
G^0_{\omega_n}({\mib r},{\mib r}') 
- \sum_{{\mibs r}_1,{\mibs r}_1'} 
G^0_{\omega_n}({\mib r},{\mib r}_1) 
\Delta({\mib r}_1,{\mib r}_1')
F^{\dagger}_{\omega_n}({\mib r}_1',{\mib r}'),
\label{eq:a.2}
\end{equation}
\begin{equation}
F^{\dagger}_{\omega_n}({\mib r},{\mib r}')
=\sum_{{\mibs r}_1,{\mibs r}_1'}
G^0_{-\omega_n}({\mib r}_1,{\mib r})
\Delta^\ast ({\mib r}_1,{\mib r}_1') 
G_{\omega_n}({\mib r}_1',{\mib r}') ,
\label{eq:a.3}
\end{equation}
by using the Green function of the normal state, 
\begin{equation}
G^0_{\omega_n}({\mib r},{\mib r}')
=\sum_{\mibs p} \frac{1}{i\omega_n-\varepsilon_{\mibs p}} 
e^{i{\mibs p}\cdot({\mib r} - {\mib r}')}. 
\label{eq:a.4}
\end{equation}

For convenience, we introduce the center of mass coordinate 
$\tilde{\mib r} = ( {\mib r}+{\mib r}' )/2$ and the relative coordinate 
$\tilde{\tilde{\mib r}} = {\mib r}-{\mib r}'$ of the Cooper pair. 
Then, we have ${\mib r}= \tilde{\mib r} + \tilde{\tilde{\mib r}}/2 $, 
${\mib r}'= \tilde{\mib r} - \tilde{\tilde{\mib r}}/2 $ and 
${\mib \nabla}_{{\mibs r}} - {\mib \nabla}_{{\mibs r}'}
= 2{\mib \nabla}_{\tilde{\tilde{\mibs r}}} $. 
The pair potential and the pairing interaction are, respectively, 
assumed to be 
\begin{equation}
\Delta(\tilde{\mib r},{\mib k}) \equiv 
 \sum_{\tilde{\tilde{\mibs r}}}e^{-i {\mibs k} \cdot \tilde{\tilde{\mibs r}}} 
\Delta({\mib r},{\mib r}')
= \eta(\tilde{\mib r}) \phi({\mib k}) , 
\label{eq:a.5}
\end{equation}
\begin{equation}
V({\mib k}-{\mib k}') \equiv
 \sum_{\tilde{\tilde{\mibs r}}}e^{-i ({\mibs k}-{\mibs k}') \cdot 
\tilde{\tilde{\mibs r}}}V(\tilde{\tilde{\mibs r}}) 
= \bar{V} \phi^\ast({\mib k}) \phi({\mib k}') ,
\label{eq:a.6}
\end{equation}
where $\phi({\mib k})$ is a symmetry function, for example, 
$\phi({\mib k})= \sqrt{2}\cos 2 \theta$ in the $d_{x^2-y^2}$-wave pairing 
case and $\phi({\mib k})=1$ in the isotropic $s$-wave pairing case. 
Then, the Fourier transformation of $\Delta({\mib r},{\mib r}')$ is given by 
\begin{equation}
\Delta({\mib r},{\mib r}')
=\sum_{{\mibs q},{\mibs k}} e^{i {\mibs q} \cdot \tilde{\mibs r} 
+ i {\mibs k} \cdot \tilde{\tilde{\mibs r}} } 
\eta({\mib q}) \phi({\mib k})  .
\label{eq:a.7}
\end{equation}
From Eqs. (\ref{eq:a.5}) and (\ref{eq:a.6}), the self-consistent condition 
(\ref{eq:a.1}) is rewritten as 
\begin{equation}
\eta^\ast (\tilde{\mibs r}) = \bar{V} T\sum_{\omega_n}\sum_{{\mibs k}} 
\phi({\mib k})
\sum_{\tilde{\tilde{\mibs r}}}e^{i {\mibs k} \cdot \tilde{\tilde{\mibs r}}}
F^{\dagger}_{\omega_n}({\mib r},{\mib r}') .
\label{eq:a.8}
\end{equation}
On the other hand, the current density is given by 
\begin{equation}
{\mib j}({\mib r})
=\frac{i|e|}{m}
\left({\mib \nabla}_{{\mibs r}} - {\mib \nabla}_{{\mibs r}'} \right)
T\sum_{\omega_n}G_{\omega_n}({\mib r},{\mib r}')\bigg|_{{\mibs r}={\mibs r}'}  
=\frac{2i|e|}{m} T\sum_{\omega_n} {\mib \nabla}_{\tilde{\tilde{\mibs r}}} 
G_{\omega_n}({\mib r},{\mib r}') \bigg|_{\tilde{\tilde{\mibs r}} =0}  .
\label{eq:a.9}
\end{equation}
From Eqs. (\ref{eq:a.2}) and  (\ref{eq:a.3}), 
$ G_{\omega_n}({\mib r},{\mib r}') $ in Eq. (\ref{eq:a.9}) and 
$F^{\dagger}_{\omega_n}({\mib r},{\mib r}')$ in Eq. (\ref{eq:a.8}) are 
expanded in the power of $\Delta$ as follows,  
\begin{equation}
G_{\omega_n}({\mib r},{\mib r}')=
G^0_{\omega_n}({\mib r},{\mib r}')
+G^{(2)}_{\omega_n}({\mib r},{\mib r}')
+G^{(4)}_{\omega_n}({\mib r},{\mib r}') + \cdots ,
\label{eq:a.10}
\end{equation}
\begin{equation}
F^{\dagger}_{\omega_n}({\mib r},{\mib r}')=
F^{\dagger (1)}_{\omega_n}({\mib r},{\mib r}')
+F^{\dagger (3)}_{\omega_n}({\mib r},{\mib r}')
+F^{\dagger (5)}_{\omega_n}({\mib r},{\mib r}') + \cdots , 
\label{eq:a.11}
\end{equation}
where 
\begin{eqnarray}
G^{(2)}_{\omega_n}({\mib r},{\mib r}')=
&&
-\sum_{{\mibs r}_1,{\mibs r}_2,{\mibs r}_1',{\mibs r}_2'}
G^0_{\omega_n}  ({\mib r}, {\mib r}_1 )
G^0_{-\omega_n} ({\mib r}_2, {\mib r}_1')
G^0_{\omega_n}  ({\mib r}_2',{\mib r}')
\Delta     ({\mib r}_1, {\mib r}_1')
\Delta^\ast({\mib r}_2, {\mib r}_2'), 
\label{eq:a.12} \\
G^{(4)}_{\omega_n}({\mib r},{\mib r}')=
&&
\sum_{{\mibs r}_1 \sim {\mibs r}_4,{\mibs r}_1' \sim {\mibs r}_4'}
G^0_{\omega_n}  ({\mib r}, {\mib r}_1 )
G^0_{-\omega_n} ({\mib r}_2, {\mib r}_1')
G^0_{\omega_n}  ({\mib r}_2',{\mib r}_3 )
G^0_{-\omega_n} ({\mib r}_4, {\mib r}_3')
G^0_{\omega_n}  ({\mib r}_4',{\mib r}') 
\nonumber \\
&& \hspace{3cm} \times
\Delta     ({\mib r}_1, {\mib r}_1')
\Delta^\ast({\mib r}_2, {\mib r}_2') 
\Delta     ({\mib r}_3, {\mib r}_3')
\Delta^\ast({\mib r}_4, {\mib r}_4'),
\label{eq:a.13} \\
F^{\dagger (1)}_{\omega_n}({\mib r},{\mib r}')=
&&
\sum_{{\mibs r}_1,{\mibs r}_1'}
G^0_{-\omega_n}({\mib r}_1,{\mib r})
G^0_{\omega_n}({\mib r}_1',{\mib r}')  
\Delta^\ast({\mib r}_1,{\mib r}_1') , 
\label{eq:a.14} \\
F^{\dagger (3)}_{\omega_n}({\mib r},{\mib r}')= 
&&
-\sum_{{\mibs r}_1 \sim {\mibs r}_3,{\mibs r}_1' \sim {\mibs r}_3'}
G^0_{-\omega_n}({\mib r}_1 ,{\mib r} )
G^0_{\omega_n} ({\mib r}_1',{\mib r}_2 ) 
G^0_{-\omega_n}({\mib r}_3 ,{\mib r}_2')
G^0_{\omega_n} ({\mib r}_3',{\mib r}')
\nonumber \\
&& \hspace{3cm} \times
\Delta^\ast({\mib r}_1,{\mib r}_1')
\Delta     ({\mib r}_2,{\mib r}_2')
\Delta^\ast({\mib r}_3,{\mib r}_3'), 
\label{eq:a.15} \\
F^{\dagger (5)}_{\omega_n}({\mib r},{\mib r}')= 
&&
\sum_{{\mibs r}_1 \sim {\mibs r}_5,{\mibs r}_1' \sim {\mibs r}_5'}
G^0_{-\omega_n}({\mib r}_1 ,{\mib r} )
G^0_{\omega_n} ({\mib r}_1',{\mib r}_2 )
G^0_{-\omega_n}({\mib r}_3 ,{\mib r}_2')
G^0_{\omega_n} ({\mib r}_3',{\mib r}_4 )
G^0_{-\omega_n}({\mib r}_5 ,{\mib r}_4')
G^0_{\omega_n} ({\mib r}_5',{\mib r}' )
\nonumber \\ 
&& \hspace{3cm} \times
\Delta^\ast({\mib r}_1,{\mib r}_1')
\Delta     ({\mib r}_2,{\mib r}_2')
\Delta^\ast({\mib r}_3,{\mib r}_3')
\Delta     ({\mib r}_4,{\mib r}_4')
\Delta^\ast({\mib r}_5,{\mib r}_5') . 
\label{eq:a.16}
\end{eqnarray}

We substitute Eqs. (\ref{eq:a.4}) and (\ref{eq:a.7}) into Eqs. 
(\ref{eq:a.12}) - (\ref{eq:a.16}), and perform the integrals in 
Eqs. (\ref{eq:a.8}) - (\ref{eq:a.16}). 
Since ${\mib q}$ corresponds to the spatial variation of the pair potential 
($|{\mib q}| \sim 1/\xi$) and ${\mib k}$ the relative motion of the Cooper 
pair ($|{\mib k}| \sim k_F$), that is, $|{\mib q}| \ll |{\mib k}|$, 
we can take 
\begin{equation}
\varepsilon_{{\mibs k}+{\mibs q}/2} \simeq \varepsilon_k 
+ \frac{v_F}{2} \hat{\mib k}\cdot{\mib q}, \qquad 
\phi({\mib k}+{\mib q}) \simeq \phi({\mib k}), 
\label{eq:a.17}
\end{equation}
where $\hat{\mib k}= {\mib k}/|{\mib k}| $, and 
Eqs. (\ref{eq:a.8}) - (\ref{eq:a.16}) are calculated by expanded also  
in the power of $\hat{\mib k}\cdot{\mib q}$. 
The ${\mib k}$-integral is treated as 
$\sum_{\mibs k}=N_F \int d \varepsilon_k \int d \theta /2 \pi $. 
We write $\int d \theta /2 \pi ( \cdots )= \langle \cdots 
\rangle_{\hat{\mibs{k}}} $ for simplicity.

As a result, the GL equation (\ref{eq:a.8}) and the current density 
(\ref{eq:a.9}) are written as 

\begin{eqnarray}
\eta^\ast(r) =
&&
\sum_{\mibs q} \bar{V} N_F \sum^\infty_{n=0} 
A_n \Bigl\langle (- {\ooalign{\hfil/\hfil\crcr$q$}} )^n 
|\phi({\mib k})|^2 \Bigr\rangle_{\hat{\mibs{k}}} \eta^\ast({\mib q}) 
e^{-i {\mibs q}\cdot {\mibs r}}
\nonumber \\
&&
+ \sum_{{\mibs q}_1 \sim {\mibs q}_3} \bar{V} N_F 
\sum^\infty_{{\scriptstyle n_1 \sim n_4=0} \atop 
{\scriptstyle (n=n_1+n_2+n_3+n_4)}} 
\Bigl\langle \left\{ 2A_{n+2} 
+ ( {\ooalign{\hfil/\hfil\crcr$q$}}_1 
+   {\ooalign{\hfil/\hfil\crcr$q$}}_3 ) A_{n+3} \right\} 
(-{\ooalign{\hfil/\hfil\crcr$q$}}_1)^{n_1}
(-{\ooalign{\hfil/\hfil\crcr$q$}}_2)^{n_2}
(-{\ooalign{\hfil/\hfil\crcr$q$}}_3)^{n_3}
(-{\ooalign{\hfil/\hfil\crcr$q$}}_4)^{n_4} 
|\phi({\mib k})|^4 \Bigr\rangle_{\hat{\mibs{k}}} 
\nonumber \\
&& \hspace{2cm}\times 
\eta^\ast({\mib q}_1) \eta({\mib q}_2) \eta^\ast ({\mib q}_3)
e^{i(-{\mibs q}_1 + {\mibs q}_2 - {\mibs q}_3) \cdot {\mib r}}
\nonumber \\
&&
+ \sum_{{\mibs q}_1\sim {\mibs q}_5} 6 \bar{V} N_F 
A_4 \Bigl\langle |\phi({\mib k})|^6 \Bigr\rangle_{\hat{\mibs{k}}}
\eta^\ast({\mib q}_1) \eta({\mib q}_2) \eta^\ast({\mib q}_3)
\eta({\mib q}_4) \eta^\ast({\mib q}_5)
e^{i(-{\mibs q}_1+{\mibs q}_2-{\mibs q}_3+{\mibs q}_4-{\mibs q}_5)
\cdot {\mib r}} ,
\label{eq:a.18}
\end{eqnarray}
\begin{eqnarray}
{\mib j}({\mib r}) = 
&&
\sum_{{\mibs q}_1,{\mibs q}_2} 2 |e| N_F v_F
\sum^\infty_{n_1,n_2=0} A_{n_1+n_2+1} \Bigl\langle \hat{\mib k} 
{\ooalign{\hfil/\hfil\crcr$q$}}_1^{n_1}
{\ooalign{\hfil/\hfil\crcr$q$}}_2^{n_2} 
|\phi({\mib k})|^2 \Bigr\rangle_{\hat{\mibs k}} 
\eta({\mib q}_1) \eta^\ast({\mib q}_2) 
e^{i({\mibs q}_1-{\mibs q}_2) \cdot {\mib r}}
\nonumber \\
&&
+\sum_{{\mibs q}_1\sim {\mibs q}_4} 4 |e| N_F v_F A_4 \Bigl\langle \hat{\mib k}
( 2 {\ooalign{\hfil/\hfil\crcr$q$}}_1 
+  {\ooalign{\hfil/\hfil\crcr$q$}}_2 
+  {\ooalign{\hfil/\hfil\crcr$q$}}_3 
+2  {\ooalign{\hfil/\hfil\crcr$q$}}_4 ) |\phi({\mib k})|^4 
\Bigr\rangle_{\hat{\mibs k}} \eta({\mib q}_1) \eta^\ast({\mib q}_2) 
\eta({\mib q}_3) \eta^\ast({\mib q}_4) 
e^{i({\mibs q}_1-{\mibs q}_2+{\mibs q}_3-{\mibs q}_4) \cdot {\mib r} }, 
\label{eq:a.19}
\end{eqnarray}
\noindent
where $\ooalign{\hfil/\hfil\crcr$q$}_i 
=v_F \hat{\mib k}\cdot {\bf {\mib q}}_i$ and 
$A_{n}$ is defined as follows:
\begin{equation}
A_n 
= \pi T\sum^{\infty}_{\omega_n=-\infty}\frac{1}{|\omega_n|(2i\omega_n)^n} 
=\left\{ \begin{array}{ll} 
{\displaystyle \ln\frac{2\omega_D\gamma}{\pi T} }, 
& \mbox{(for $n=0$)} \\
{\displaystyle \frac{2(-)^{n/2}}{(2\pi T)^{n}} 
\left( 1- \frac{1}{2^{n+1}} \right) \zeta(n+1)},   
& \mbox{(for even $n\geq 2$)} \\ 
 0,  & \mbox{(for odd $n$)} 
\end{array} \right. 
\label{eq:a.20}
\end{equation}
with the Debye frequency $\omega_D$ and the Euler constant $\gamma$.
In the second term of Eq. (\ref{eq:a.18}), we define 
${\mib q}_4 ={\mib q}_1 -{\mib q}_2 +{\mib q}_3$. 
In the last terms of Eqs. (\ref{eq:a.18}) and (\ref{eq:a.19}), 
we present only the contributions of the lowest order of ${\mib q}$.

By performing the ${\mib q}$-integral in Eqs. (\ref{eq:a.18}) and 
(\ref{eq:a.19}), ${\mib q}_i^n \eta({\mib q}_i) e^{i {\mibs q}_i \cdot 
{\mibs r}}$ are replaced by 
$(-i {\mib D}_i)^n \eta({\mib r}_i) |_{{\mib r}_i={\mib r}}$, 
where ${\mib D_i}={\mib \nabla}_{{\mibs r}_i} +i(2 \pi /\phi_0)
{\mib A}({\mib r}_i)$ is introduced instead of ${\mib \nabla}_{{\mibs r}_i}$ 
for conserving the Gauge invariance. 
Then (\ref{eq:a.18}) and (\ref{eq:a.19}) are rewritten as 
\begin{eqnarray}
\eta^\ast (r) =
&&
N_F \bar{V} \sum^\infty_{n=0} A_n  \Bigl\langle 
(-i {\ooalign{\hfil/\hfil\crcr$D$}}^\ast)^n 
|\phi({\mib k})|^2 \Bigr\rangle_{\hat{\mibs{k}}} \eta^\ast({\mib r}) 
\nonumber \\
&&
+ N_F \bar{V} \sum_{{\scriptstyle n_1 \sim n_4=0} \atop 
{\scriptstyle (n=n_1+n_2+n_3+n_4)}}^{\infty} 
\Bigl\langle \left\{ 2A_{n+2} 
+i( {\ooalign{\hfil/\hfil\crcr$D$}}_1^\ast
   +{\ooalign{\hfil/\hfil\crcr$D$}}_3^\ast) A_{n+3} \right\}
(-i\ooalign{\hfil/\hfil\crcr$D$}_1^\ast)^{n_1}
( i\ooalign{\hfil/\hfil\crcr$D$}_2     )^{n_2}
(-i\ooalign{\hfil/\hfil\crcr$D$}_3^\ast)^{n_3}
( i\ooalign{\hfil/\hfil\crcr$D$}_4     )^{n_4}
|\phi({\mib k})|^4 \Bigr\rangle_{\hat{\mibs k}} 
\nonumber \\
&& \hspace{0.5cm}\times 
\eta^\ast({\mib r}_1) \eta({\mib r}_2) \eta^\ast({\mib r}_3) 
\Bigr|_{{\mibs r}_1 ={\mibs r}_2 = {\mibs r}_3={\mibs r}} 
+  6 N_F \bar{V} A_4 \Bigl\langle |\phi({\mib k})|^6 
\Bigr\rangle_{\hat{\mibs{k}}}
|\eta({\mib r})|^4 \eta^\ast({\mib r}) , 
\label{eq:a.21}
\end{eqnarray}
\begin{eqnarray}
{\mib j}({\mib r}) = 
&&
 2 |e| N_F v_F
\sum^\infty_{n_1,n_2=0}
A_{n_1+n_2+1} \Bigl\langle \hat{\mib k} 
(-i{\ooalign{\hfil/\hfil\crcr$D$}}_1)^{n_1}
( i{\ooalign{\hfil/\hfil\crcr$D$}}_2^\ast)^{n_2} 
|\phi({\mib k})|^2 \Bigr\rangle_{\hat{\mibs k}} 
\eta({\mib r}_1) \eta^\ast({\mib r}_2)
\Bigr|_{{\mibs r}_1 ={\mibs r}_2 ={\mibs r}}
\nonumber \\
&&
+ 4 i |e| N_F v_F A_4 \Bigl\langle \hat{\mib k}
 (-2 {\ooalign{\hfil/\hfil\crcr$D$}}_1 
  +  {\ooalign{\hfil/\hfil\crcr$D$}}_2^\ast 
  -  {\ooalign{\hfil/\hfil\crcr$D$}}_3 
  +2 {\ooalign{\hfil/\hfil\crcr$D$}}_4^\ast ) |\phi({\mib k})|^4 
\Bigr\rangle_{\hat{\mibs k}} \eta({\mib r}_1) \eta^\ast({\mib r}_2) 
\eta({\mib r}_3) \eta^\ast({\mib r}_4) 
\Bigr|_{{\mibs r}_1 ={\mibs r}_2 = {\mibs r}_3={\mibs r}_4={\mibs r}} , 
\label{eq:a.22}
\end{eqnarray}
\noindent
where ${\ooalign{\hfil/\hfil\crcr$D$}} = v_F \hat{\mib k}\cdot{\mib D}$. 
In Eq. (\ref{eq:a.21}), we define 
${\mib D}_4= -{\mib D}_1^\ast -{\mib D}_2 -{\mib D}_3^\ast$.

We substitute $\phi({\mib k})= \sqrt{2} \cos 2 \theta $ 
(in the $d_{x^2-y^2}$-wave case) and $\hat{\mib k}\cdot{\mib D} = 
D_x \cos\theta +D_y \sin\theta $, and perform the $\theta$-integral. 
Then, using the relation 
$(N_F \bar{V})^{-1}=\ln (2 \omega_D \gamma / \pi T_c )$, 
we obtain Eqs. (\ref{eq:2.2}) and (\ref{eq:2.3}). 

We confirm that the GL equation (\ref{eq:2.2}) and the current expression 
(\ref{eq:2.3}) are also derived by expanding the quasi-classical 
Eilenberger equation in the powers of ${\mib D}$ and $\eta$ 
following the method presented by Schopohl and Tewordt.~\cite{Schopohl}

\section{Derivation of Eq. (\ref{eq:3.3}) }
\label{sec:ap.b}

Around a vortex, in general, $\eta({\mib r})$ of the pair potential 
$\Delta({\mib r},{\mib k})$ in Eq. (\ref{eq:2.1}) can be expanded by 
$e^{il \phi}$ ($l$: integer) as follows, 
\begin{equation}
\eta({\mib r})= \sum_l f_l(r) e^{il \phi}. 
\label{eq:b.1}
\end{equation}
Then, as for the $U_{2,x}R$ and  $U_{2,x-y}e^{-i \pi /2}R$ in Eq. 
(\ref{eq:3.2}), the transformations of $\Delta({\mib r},{\mib k})$ 
are given by 
\begin{equation}
U_{2,x}R \Delta({\mib r},{\mib k}) = 
\sum_l f_l^\ast (r) e^{i l \phi} \sqrt{2} \cos 2 \theta, 
\label{eq:b.2}
\end{equation}
\begin{equation}
U_{2,x-y}e^{-i \pi /2}R \Delta({\mib r},{\mib k})
= -\sum_l f_l^\ast (r) e^{-i(l+1)\pi/2 +il \phi}
\sqrt{2} \cos 2 \theta.
\label{eq:b.3}
\end{equation}
If the symmetry $D_4(E)$ is conserved, Eqs. (\ref{eq:b.2}) and (\ref{eq:b.3}) 
should be equal to $\Delta({\mib r},{\mib k})$. 
Then, we obtain the relation, 
\begin{equation}
f_l(r)=f_l^\ast(r), \qquad 
e^{-i(l+1)\pi/2}= -1. 
\label{eq:b.4}
\end{equation}
From Eq. (\ref{eq:b.4}), $f_l(r)$ is real, and $l=4n+1$ ($n$: integer). 
Writing $f_{l=4n+1}(r)$ as $f_n(r)$, we obtain Eq. (\ref{eq:3.3}). 


\end{full}
\end{document}